\documentclass{article}

\usepackage[preprint]{neurips_2025}

\usepackage[utf8]{inputenc} 
\usepackage[T1]{fontenc} 
\usepackage{hyperref} 
\usepackage{url} 
\usepackage{booktabs} 
\usepackage{amsfonts} 
\usepackage{nicefrac} 
\usepackage{microtype} 

\usepackage{algorithm}
\usepackage{algpseudocode}
\usepackage{listings}

\usepackage{xcolor} 
\usepackage{graphicx}
\usepackage{adjustbox}
\usepackage{pifont}
\usepackage{amsmath}
\usepackage{graphicx}
\usepackage{pgfplots}
\usepackage{comment}
\usepackage{amsmath,amssymb} %
\usepackage{color}
\usepackage{soul}
\usepackage{svg}
\usepackage{wrapfig}
\usepackage{array}
\usepackage{xspace}
\usepackage{subcaption}
\usepackage{wrapfig}
\usepackage{mathrsfs}
\usepackage{sidecap}
\usepackage{colortbl} 

\usepackage{bm}
\usepackage{enumitem}
\usepackage{xspace}
\usepackage{cutwin}
\definecolor{custompink}{RGB}{255, 20, 147} 

\hypersetup{
    colorlinks=true, 
    linkcolor=custompink, 
    filecolor=custompink,
    urlcolor=custompink
}
\usepackage{multirow}
\usepackage{graphicx}
\usepackage{wrapfig}
\usepackage{float}

\newcommand{\cmark}{\ding{51}}

\DeclareMathOperator{\Pool}{\mathcal{P}}
\DeclareMathOperator{\Voxel}{\mathbf{V}}
\DeclareMathOperator{\Conv}{\text{Conv}}
\DeclareMathOperator{\Defconv}{\text{DefConv}}

\newcommand{\SecondBest}[1]{\textcolor{black}{\underline{{#1}}}}
\newcommand{\Best}[1]{\textcolor{black}{\textbf{{#1}}}}
\newcommand{\highlight}[1]{\textcolor{black}{\emph{\textbf{#1}}}}

\newcommand{\Method}{VoxDet\xspace}
\newcommand{\Trick}{VoxNT\xspace}
\newcolumntype{C}[1]{>{\centering\arraybackslash}p{#1}} 

\definecolor{White}{rgb}{1.,0.,1.}
\definecolor{first}{rgb}{.8,.0,.0}
\definecolor{second}{rgb}{.0,.6,.0}
\definecolor{third}{rgb}{.0,.0,.8}

\definecolor{ceiling}{RGB}{214,  38, 40}
\definecolor{floor}{RGB}{43, 160, 4}
\definecolor{wall}{RGB}{158, 216, 229}
\definecolor{window}{RGB}{114, 158, 206}
\definecolor{chair}{RGB}{204, 204, 91}
\definecolor{bed}{RGB}{255, 186, 119}
\definecolor{sofa}{RGB}{147, 102, 188}
\definecolor{table}{RGB}{30, 119, 181}
\definecolor{tvs}{RGB}{160, 188, 33}
\definecolor{furniture}{RGB}{255, 127, 12}
\definecolor{objects}{RGB}{196, 175, 214}

\definecolor{car}{rgb}{0.39215686, 0.58823529, 0.96078431}
\definecolor{bicycle}{rgb}{0.39215686, 0.90196078, 0.96078431}
\definecolor{motorcycle}{rgb}{0.11764706, 0.23529412, 0.58823529}
\definecolor{truck}{rgb}{0.31372549, 0.11764706, 0.70588235}
\definecolor{othervehicle}{rgb}{0.39215686, 0.31372549, 0.98039216}
\definecolor{person}{rgb}{1.        , 0.11764706, 0.11764706}
\definecolor{bicyclist}{rgb}{1.        , 0.15686275, 0.78431373}
\definecolor{motorcyclist}{rgb}{0.58823529, 0.11764706, 0.35294118}
\definecolor{road}{rgb}{1.        , 0.        , 1.        }
\definecolor{parking}{rgb}{1.        , 0.58823529, 1.        }
\definecolor{sidewalk}{rgb}{0.29411765, 0.        , 0.29411765}
\definecolor{otherground}{rgb}{0.68627451, 0.        , 0.29411765}
\definecolor{building}{rgb}{1.        , 0.78431373, 0.        }
\definecolor{fence}{rgb}{1.        , 0.47058824, 0.19607843}
\definecolor{vegetation}{rgb}{0.        , 0.68627451, 0.        }
\definecolor{trunk}{rgb}{0.52941176, 0.23529412, 0.        }
\definecolor{terrain}{rgb}{0.58823529, 0.94117647, 0.31372549}
\definecolor{pole}{rgb}{1.        , 0.94117647, 0.58823529}
\definecolor{trafficsign}{rgb}{1.        , 0.        , 0.        }
\definecolor{otherstructure}{rgb}{0.98039215, 0.58823529, 0.}
\definecolor{otherobject}{rgb}{0.19607843, 1.        , 1.        }

\makeatletter
\newcommand{\car@semkitfreq}{3.92}
\newcommand{\bicycle@semkitfreq}{0.03}
\newcommand{\motorcycle@semkitfreq}{0.03}
\newcommand{\truck@semkitfreq}{0.16}
\newcommand{\othervehicle@semkitfreq}{0.20}
\newcommand{\person@semkitfreq}{0.07}
\newcommand{\bicyclist@semkitfreq}{0.07}
\newcommand{\motorcyclist@semkitfreq}{0.05}
\newcommand{\road@semkitfreq}{15.30}
\newcommand{\parking@semkitfreq}{1.12}
\newcommand{\sidewalk@semkitfreq}{11.13}
\newcommand{\otherground@semkitfreq}{0.56}
\newcommand{\building@semkitfreq}{14.1}
\newcommand{\fence@semkitfreq}{3.90}
\newcommand{\vegetation@semkitfreq}{39.3}
\newcommand{\trunk@semkitfreq}{0.51}
\newcommand{\terrain@semkitfreq}{9.17}
\newcommand{\pole@semkitfreq}{0.29}
\newcommand{\trafficsign@semkitfreq}{0.08}
\newcommand{\semkitfreq}[1]{{\csname #1@semkitfreq\endcsname}}

\newcommand{\car@sscbkitfreq}{2.85}
\newcommand{\bicycle@sscbkitfreq}{0.01}
\newcommand{\motorcycle@sscbkitfreq}{0.01}
\newcommand{\truck@sscbkitfreq}{0.16}
\newcommand{\othervehicle@sscbkitfreq}{5.75}
\newcommand{\person@sscbkitfreq}{0.02}
\newcommand{\road@sscbkitfreq}{14.98}
\newcommand{\parking@sscbkitfreq}{2.31}
\newcommand{\sidewalk@sscbkitfreq}{6.43}
\newcommand{\otherground@sscbkitfreq}{2.05}
\newcommand{\building@sscbkitfreq}{15.67}
\newcommand{\fence@sscbkitfreq}{0.96}
\newcommand{\vegetation@sscbkitfreq}{41.99}
\newcommand{\terrain@sscbkitfreq}{7.10}
\newcommand{\pole@sscbkitfreq}{0.22}
\newcommand{\trafficsign@sscbkitfreq}{0.06}
\newcommand{\otherstructure@sscbkitfreq}{4.33}
\newcommand{\otherobject@sscbkitfreq}{0.28}
\newcommand{\sscbkitfreq}[1]{{\csname #1@sscbkitfreq\endcsname}}

\title{
  \centering
  \hspace{0.2cm}
  \raisebox{0.2ex}{\includegraphics[scale=0.4, bb=70 80 30 0]{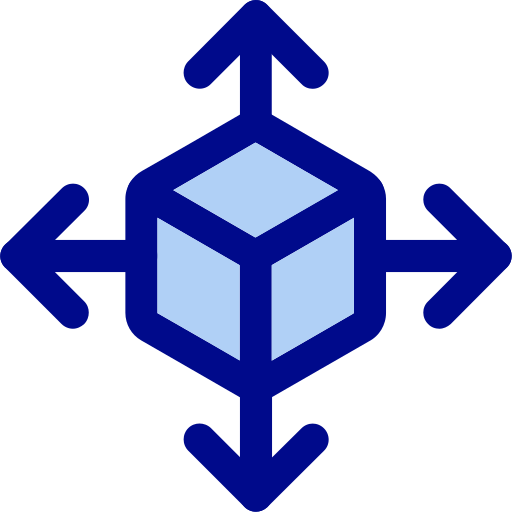}} \quad \quad
 VoxDet: Rethinking 3D Semantic Occupancy Prediction as Dense Object Detection
}

\author{
	Wuyang Li$^{1}$ \hspace{1em}
	Zhu Yu$^{2}$ \hspace{1em}
	Alexandre Alahi$^{1}$
	\\[2pt]
	$^1$École Polytechnique Fédérale de Lausanne (EPFL)\hspace{2em}
	$^2$Zhejiang University\\
         \texttt{wuyang.li@epfl.ch}
    \hspace{1em}  \texttt{yu\_zhu@zju.edu.cn} \hspace{1em}  \texttt{alexandre.alahi@epfl.ch} 
	\\\\
	Project Page: \href{https://vita-epfl.github.io/VoxDet/}{https://vita-epfl.github.io/VoxDet/}
}

\begin{document}

\maketitle

\begin{abstract} 
3D semantic occupancy prediction aims to reconstruct the 3D geometry and semantics of the surrounding environment. With dense voxel labels, prior works typically formulate it as a \emph{dense segmentation task}, independently classifying each voxel. However, this paradigm neglects critical instance-centric discriminability, leading to instance-level incompleteness and adjacent ambiguities. To address this, we highlight a "free lunch" of occupancy labels: the voxel-level class label implicitly provides insight at the instance level, which is overlooked by the community. Motivated by this observation, we first introduce a training-free \textbf{Voxel-to-Instance (\Trick) trick}: a simple yet effective method that freely converts voxel-level class labels into instance-level offset labels. Building on this, we further propose \textbf{\Method}, an instance-centric framework that reformulates the voxel-level occupancy prediction as \emph{dense object detection} by decoupling it into two sub-tasks: offset regression and semantic prediction. Specifically, based on the lifted 3D volume, \Method first uses (a) Spatially-decoupled Voxel Encoder to generate disentangled feature volumes for the two sub-tasks, which learn task-specific spatial deformation in the densely projected tri-perceptive space. Then, we deploy (b) Task-decoupled Dense Predictor to address this task via dense detection. Here, we first regress a 4D offset field to estimate distances (6 directions) between voxels and the corresponding object boundaries in the voxel space. The regressed offsets are then used to guide the instance-level aggregation in the classification branch, achieving instance-aware prediction. Experiments show that VoxDet can be deployed on both camera and LiDAR input, jointly achieving state-of-the-art results on both benchmarks. VoxDet is not only highly efficient, but also gives 63.0 IoU on the SemanticKITTI test set, \textbf{ranking 1$^{st}$} on the online leaderboard.


\end{abstract}

\section{Introduction}

Spatial AI~\cite{duan2025worldscore} is crucial for autonomous systems to perceive and interpret the complex physical world. As a critical step, precise reconstruction of geometric structures and semantics lays the foundation for scene understanding, underpinning the downstream forecasting and planning~\cite{alahi2016social}. This capability is indispensable for applications such as autonomous driving and robotic navigation.

To this end, 3D semantic occupancy prediction, derived from semantic scene completion~\cite{roldao20223d}, has attracted significant attention by simultaneously inferring complete 3D geometry and semantics through voxel representation. Prior works can be broadly categorized into LiDAR-based~\cite{SSCNet,LMSCNet,xia2023scpnet} and camera-based approaches~\cite{MonoOcc,VoxFormer,yucontext,Symphonize}. The former uses sparse 3D inputs (e.g., point clouds) to provide precise geometric information. In contrast, camera-based methods have recently demonstrated promising potential due to their flexibility and computational efficiency. They employ dedicated vision-centric algorithms to lift 2D imagery into 3D space, including Features Line-of-Sight Projection (FLoSP)~\cite{MonoScene}, depth-driven back projection~\cite{VoxFormer,Symphonize,yu2023aggregating}, and Lift-Splat-Shoot (LSS)~\cite{lss} based voxels~\cite{yucontext,li2024hierarchical,bae2024three}. These methods offer unique advantages in resource-constrained settings such as robotics, owing to their low cost, high flexibility, and real-time processing capabilities.

\begin{figure*}[t]
	\centering\includegraphics[width=1.0\linewidth]{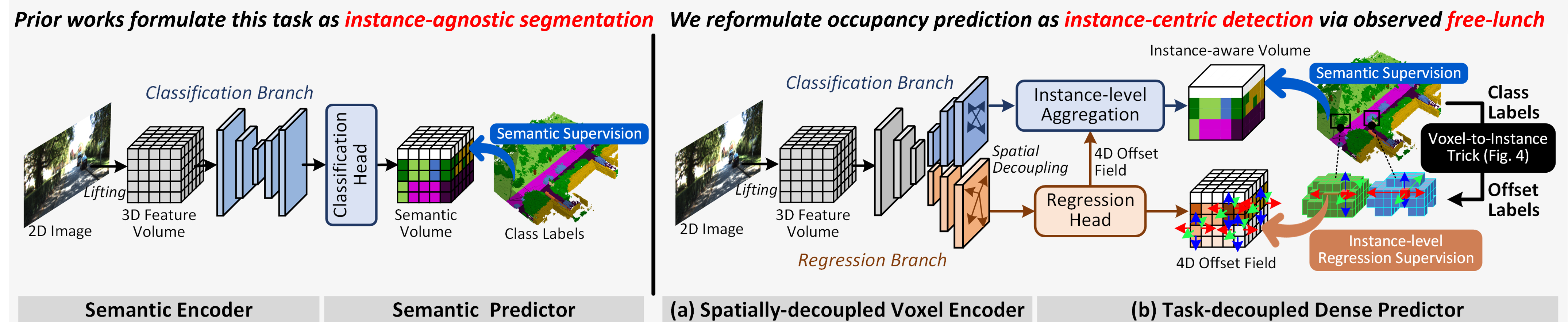}
	\caption{Schematic comparison of previous paradigm~\cite{MonoScene,bae2024three,yucontext} and our \Method. \textbf{Left:} Prior works use a 3D semantic segmentation formulation, directly predicting voxel labels agnostic to object instances. \textbf{Right:} \Method reformulates this task as dense object detection to explicitly learn with instances, which decouples it into instance-aware regression and classification. This is achieved via a Voxel-to-Instance (VoxNT) trick (Fig.~\ref{fig:trick}), inspired by our observed free-lunch in labels (Sec.~\ref{sec:motivation}).}
	\label{fig:intro}
	\vspace{-4mm}
\end{figure*}

While achieving great progress, existing methods uniformly formulate occupancy prediction as a semantic segmentation task\footnote{For simplicity, we uniformly refer to voxel recognition and completion~\cite{MonoScene} as dense classification.} on dense voxels, which fail to understand the instance-level semantics and geometry. This paradigm typically employs 3D encoders~\cite{MonoOcc,yucontext} to encode semantic patterns from the lifted volumes (see Fig.~\ref{fig:intro} \textbf{Left}), subsequently classifying each voxel independently. Due to the absent instance labels, this voxel-centric paradigm may seem viable, but it has to face severe instance-level incompleteness and adjacent ambiguities (see Fig.~\ref{fig:show}). Although prior works, such as Symphonize~\cite{Symphonize}, have noticed this issue and attempted to mitigate it with object queries from 2D images, they still optimize the 3D space via dense segmentation on each voxel. The substantial gap between the image and voxel domains prevents effective instance-driven learning, making the segmentation-based formulation fail to infer the complex environment with a lot of dynamic agents.

To address this issue, we first highlight a ``free-lunch'' of the occupancy labels: \highlight{the voxel-level class label has implicitly told the instance-level insight, which is ever-overlooked by the community.} {First, unlike in 2D images where occlusion causes overlapping objects to conflate into a single entity, instances in 3D voxels are inherently separable.} Every voxel is deterministically assigned to a single class without occlusion-induced ambiguity, making the instance-level discovery essentially realistic. {Second, instance boundaries in 3D voxel space are naturally tractable and regressible,} ensuring clear distinctions between instances. More details can be found in Sec.~\ref{sec:motivation}.

Inspired by these observations, we first develop a \textbf{Voxel-to-Instance (\Trick) Trick} (Fig.~\ref{fig:trick}) that can freely convert the voxel-level class labels to the instance-level offset labels, fully harnessing the free-lunch (Sec.~\ref{sec:motivation}). Here, the offset is defined as the Euclidean distance between each voxel and its associated instance borders. Then, these \emph{free offset labels} prompt us to rethink the segmentation-based formulation. In response to this, we propose \textbf{\Method } to reformulate \underline{Vox}el-level occupancy prediction as instance-centric object \underline{Det}ection (see Fig.~\ref{fig:intro} \textbf{Right}). Unlike prior works, \Method decouples this task into two sub-tasks: offset regression and semantic prediction. We achieve this with (a) Spatially-decoupled Voxel Encoder (SVE) and (b) Task-decoupled Dense Predictor (TDP) at the representation and prediction levels, respectively: Given 3D volumes lifted from 2D images~\cite{lss}, SVE first learns task-specific features by spatially deforming in the tri-perceptive space, which are sent to the two branches of TDP, respectively. Within TDP, we regress each voxel to the associated instance boundary by predicting a novel 4D offset field, which guides an instance-level aggregation for instance-centric occupancy prediction. 

Hence, with our new detection-based formulation, \Method achieves true instance-centric perception solely using voxel-level labels, which can be effortlessly extended to the LiDAR settings. \Method is not only highly efficient by reducing 57.9\% parameters with 1.3 $\times$ speed-up, but also achieves state-of-the-art results on both camera and LiDAR benchmarks. Notably, our method gives 63.0 IoU, \textbf{ranking 1st} on the CodaLab leaderboard, without using extra labels/data/temporal information/models.

In summary, our contributions lie in the following aspects:

\begin{itemize}
 \item By analyzing the difference of 2D pixels and 3D voxels (Sec.~\ref{sec:motivation}), we reveal the overlooked free lunch of occupancy labels for instance-level learning. With this observation, we propose a \Trick trick to freely convert voxel-level labels to instance-level offsets. As a byproduct, this trick can also identify wrong labels of dynamic objects (see Appendix and Fig.~\ref{fig:show}).
 
 \item We reformulate semantic occupancy prediction as a dense object detection task by advancing {\Method}, decoupling into two sub-tasks: offset regression and semantic prediction, for instance-level perception.

 \item We design a Spatially-decoupled Voxel Encoder (SVE) to decouple 3D volumes for our new formulation, which learns task-specific spatial deformation in the densely projected tri-perceptive space for the two sub-tasks, avoiding the misalignment between them.
 
 \item We propose a Task-decoupled Dense Predictor (TDP) to enable instance-driven prediction. This comprises a regression branch that predicts a 4D offset field, delineating instance boundaries, and a classification branch using learned offsets for instance-aware aggregation. 

\end{itemize}

\section{Related Work}

\noindent\textbf{3D Semantic Occupancy Prediction}, derived from Semantic Scene Completion~\cite{SSCNet,cao2024pasco,chambon2025gaussrender}, aims to jointly reconstruct the semantics and geometry of a surrounding environment with voxelization. Existing studies can be generally divided into LiDAR-based and camera-based methods. The former utilizes point clouds to achieve high accuracy with precise depth, which is limited by the computational cost. Camera-based methods rely solely on 2D visual inputs to generate 3D scene understanding. With the advancement of monocular vision like LSS~\cite{lss}, these approaches offer great advantages in efficiency. MonoScene~\cite{MonoScene} pioneered the camera-based setting by connecting 2D images with the 3D voxel space via the FLoSP. VoxFormer~\cite{VoxFormer} uses 3D-to-2D back-projection and disseminates the semantics of the visible queries across the entire 3D volume via MAE~\cite{MAE}. Subsequent works have further enhanced voxel representations using techniques such as tri-perceptive enhancement~\cite{yucontext,bae2024three}, diffusion models~\cite{liang2025skip}, vision-language models~\cite{wang2025vlscene,yu2024language}, geometric depth~\cite{PointDC,yan2022rignet,yan2024tri,yan2025rignet++}, and extra modalities~\cite{guo2025sgformer,wang2025l2cocc,yan2025event}, boosting downstream applications~\cite{liu2025x,li2025instantsplamp,liu2024lgs,yang2025concealgs}. However, existing methods uniformly treat this task as dense segmentation, lacking explicit instance-level perception. In contrast, we reformulate it as a dense object detection task with explicit instance-level awareness.

\noindent\textbf{Dense Object Detection}, such as point-based FCOS~\cite{tian2019fcos,wang2021fcos3d}, is a fully convolutional paradigm known for its lightweight design, efficiency, and performance, which garnered significant attention prior to the DETR series~\cite{DETR,DETR3D,DeformableDetr,rtDETR}. The core insight is the notion of ``\highlight{densely detecting like segmentation}'': every pixel within a bounding box regresses its distances in four directions (up, down, left, right) while simultaneously predicting the instance-level class label, which is a dense process like the per-pixel segmentation. The following works focus on improving this paradigm from different aspects, including considering better label assignment~\cite{zhang2020bridging,zhang2021varifocalnet} like ATSS~\cite{zhang2020bridging}, architecture search~\cite{wang2020fcos}, spatially task decoupling~\cite{chen2021disentangle}, border enhancement~\cite{qiu2020borderdet}, dense feature distillation~\cite{zheng2022localization,yang2022prediction,li2022sigma,li2022scan}, optimization signals~\cite{xu2022revisiting,lin2017focal}. Besides, dense detection has been extended to 3D vision. For instance, FCOS-3D~\cite{wang2021fcos3d} predicts 3D targets in the 2D space by regressing 3D attributes for each pixel, and UVTR~\cite{li2022unifying} predicts a 3D position for each pixel to enhance instance-level localization. Unlike previous works, we reverse the philosophy via a ``\highlight{segmenting like dense detection}'' framework that endows voxel-based segmentation with instance-level awareness, thereby eliminating semantic and geometric ambiguity in occupancy prediction. Notably, our approach not only lifts the pixel-based regression into the 3D voxel space but also eliminates the need for instance-level bounding box labels.

\section{Preliminaries and Motivation \label{sec:motivation}} 

We start by analyzing the differences between 2D pixel and 3D voxel space with respect to semantic-level classification and instance-level regression, from the perspective of dense perception~\cite{tian2019fcos}. We then clarify our observations and insights on the ever-overlooked ``free-lunch” in voxel-level class labels and explain the associated motivation for the following methodology.

\begin{SCfigure}[][t]
\centering
\includegraphics[width=0.6\linewidth]{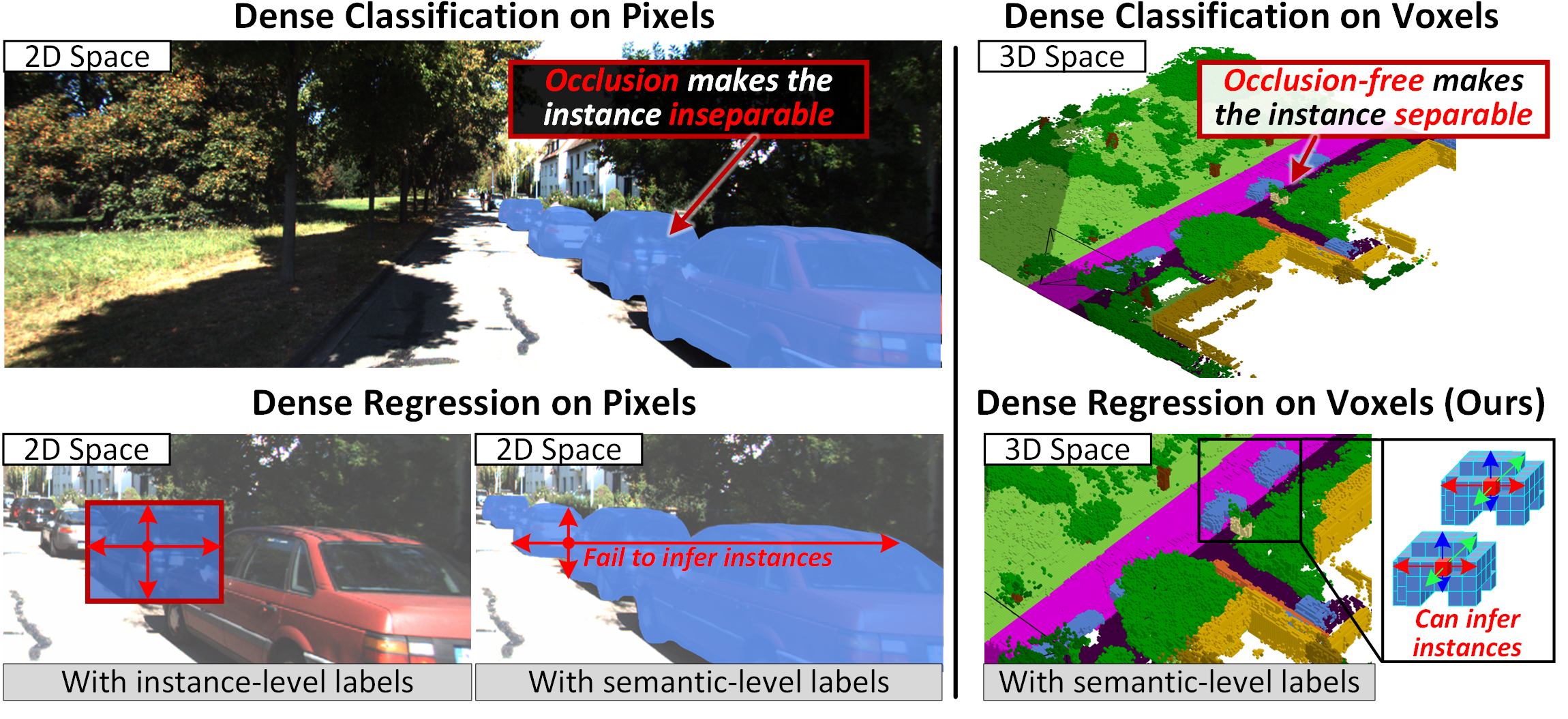}
\caption{Conceptual comparison between 2D pixels and 3D voxels regarding semantic-level dense classification and instance-level regression. \textbf{Left:} 2D occlusion makes instance perception fail with semantic labels. \textbf{Right:} The occlusion-free nature of 3D space enables instances separable only using semantic labels.\vspace{-5pt} }
\label{fig:insight}
\vspace{-10pt}
\end{SCfigure}

\noindent\textbf{Semantic-based Dense Classification.} In Fig.~\ref{fig:insight} (\textbf{Top}), it can be observed that occlusion in the 2D space often leads to the merging of distinct object instances (e.g., the blue-highlighted cars) into a single entity (\textbf{Left}). Due to the perceptive projection~\cite{faugeras1993three}, multiple 3D points at varying depths in the world coordinate system converge into a single pixel of the image. In contrast, the 3D voxel space inherently avoids most of such occlusion, as each voxel is uniquely assigned a class label, including the \emph{empty} class (\textbf{Right}). \emph{Consequently, the occlusion-free property of 3D voxels facilitates the natural separation of object instances}, although minor ambiguities may arise in cases such as the densely overlapping foliage of adjacent trees, which typically do not affect the scene understanding.

\noindent\textbf{Instance-based Dense Regression.} In Fig.~\ref{fig:insight} (\textbf{Bottom}), each pixel and voxel aims to regress the distance to the instance border~\cite{tian2019fcos}. In the 2D domain (\textbf{Left}), such regression supports instance-level perception via bounding box labels; however, it fails to infer instance cues with semantic labels. \emph{Conversely, in the 3D voxel space, the natural separability of instances makes it possible to regress instances without instance labels} (\textbf{Right}). For example, consider the red voxel inside the highlighted car in the figure. As the voxel space is occlusion‐free, we can regress from this voxel to its instance boundary by identifying adjacent voxels with differing semantic labels (see Fig.~\ref{fig:trick}). This property enables us to infer instances using only semantic annotations, a valuable yet overlooked \emph{free lunch}.

\noindent\textbf{Motivation and Insight}. Inspired by these observations, we aim to rethink \underline{Vox}el-level semantic occupancy prediction as object \underline{Det}ection~\cite{tian2019fcos}, termed \Method, which adopts a novel ``segmenting like dense detection'' philosophy. Instead of directly recognizing each independent voxel~\cite{MonoOcc,yucontext,bae2024three,VoxFormer}, \Method decouples it into regression and classification sub-tasks in the 3D voxel space, where, in particular, the core innovative regression (or offset) branch enables explicit instance-level 3D regression. The learned offsets are subsequently used for instance-level aggregation, facilitating scene understanding. Note that this fundamentally differs from and advances beyond the traditional segmentation paradigm by \highlight{explicitly lifting voxels to instances}, with potential applicability to the broader occupancy community, such as multi-camera settings.

\section{Methodology}

\textbf{Overview.} Fig.~\ref{fig:overview} shows the overall workflow of our \Method. Given RGB input, we follow previous works~\cite{yucontext} to conduct \textbf{(a)} 2D-to-3D lifting to generate 3D feature volumes $\Voxel$, which uses the estimated depth $\mathbf{Z}$ given by the arbitrary depth estimator~\cite{Adabins,MobileStereoNet}. Then, we send $\Voxel$ to \textbf{(b)} Spatially-decoupled Voxel Encoder to generate the disentangled feature for dense classification $\Voxel_{\text{cls}}$ and regression $\Voxel_{\text{reg}}$. Here, we encourage the two tasks to learn spatially decoupled features to avoid task-misalignment~\cite{xu2022revisiting}, which is achieved in a densely projected tri-perceptive (TPV) space. Next, the decoupled volumes are sent to \textbf{(c)} Task-decoupled Dense Predictor. In this part, we first regress a 4D offset field $\Delta$ to estimate the distance between each voxel $\Voxel_{i,j,k}$ to the instance boundary $\Delta_{i,j,k}\in \mathbb{R}^6$ in six directions (see Fig.~\ref{fig:trick}). The learned offset $\Delta$ is subsequently sent to the classification branch to guide instance-level aggregation, thereby achieving instance-aware semantic occupancy prediction.

\subsection{2D-to-3D Lifting}

\begin{figure*}[t]
	\centering\includegraphics[width=1.0\linewidth]{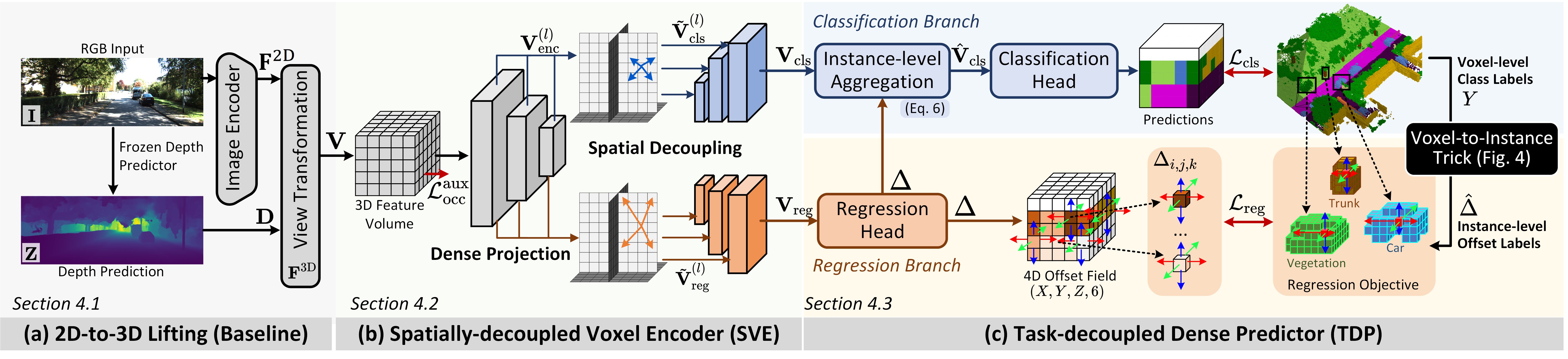}
	\caption{Overview of our \Method. After 2D-to-3D lifting, \Method spatially decouples 3D volumes $\Voxel$ into two task-specific branches, learning different spatial deformations in the densely projected tri-perceptive space. Then, \Method regresses a 4D offset field $\Delta$ towards instance boundaries with $\Voxel_{\text{reg}}$, serving for the instance-level aggregation with $\Voxel_{\text{cls}}$ in the classification branch.}
	\label{fig:overview}
\vspace{-4mm}
\end{figure*}

We follow previous works~\cite{yucontext,bae2024three,VoxFormer} to conduct the same 2D-to-3D lifting (Fig.~\ref{fig:overview}a), outputting the 3D feature volume $\Voxel\in \mathbb{R}^{X\times Y\times Z\times C}$, where ${X}$, ${Y}$, ${Z}$, and $C$ is the depth, width, height, and channel respectively. The process is briefly described as follows. More details are in the Appendix.

Given the RGB image $\mathbf{I}$, we extract the 2D image feature $\mathbf{F}^{\text{2D}}$ using the image encoder, and estimate the depth map $\mathbf{Z}$ with the frozen depth estimator~\cite{Adabins,MobileStereoNet} following the unified practice~\cite{yucontext,VoxFormer,bae2024three}. Then, we estimate the depth probability $\mathbf{D}$ using LSS~\cite{lss}. Based on these, we can establish the 3D feature $\mathbf{F}^{\text{3D}}$ using the fused depth probability $\mathbf{D}$ and pre-extracted 2D feature $\mathbf{F}^{\text{2D}}$~\cite{yucontext}, which is subsequently projected onto the voxel grid to generate the 3D volume $\Voxel$. In this procedure, each voxel in $\Voxel$ is able to query the information from 3D features $\mathbf{F}^{\text{3D}}$ via deformable cross-attention~\cite{dai2017deformable}.

\subsection{Spatially-decoupled Voxel Encoder}

Given the 3D volume $\Voxel$, we aim to extract task-specific representations for the two tasks. Prior encoders~\cite{li2025u,cciccek20163d} are designed for voxel segmentation, which lacks the spatial context for regression with task-misalignment~\cite{song2020revisiting,li2021htd,chen2021disentangle}. Hence, we spatially decouple $\Voxel$ into $\Voxel_\text{cls}$ for classification and $\Voxel_\text{reg}$ for regression (Fig.~\ref{fig:overview}b) in a densely projected TPV space. Fig.~\ref{fig:feat} shows the decoupling effect.

\noindent\textbf{Dense Projection.} Given the 3D volume $\Voxel$, we use a shared encoder to extract $L$-level task-shared feature volumes $\{{\Voxel}^{(l)}_{\text{enc}}\}_{l=1}^{L}$. Due to the task-misaligned feature preference~\cite{xu2022revisiting}, directly using the shared volume for the two tasks will severely influence the instance-level perception. To resolve this, we spatially decouple task‐specific features in a densely projected tri‐perceptive (TPV) space~\cite{TPVFormer}. This can avoid the large computational cost of decoupling in 3D, effectively reducing the dimension to 2D. To this end, a dense projection $\Pool_{{d}}(\cdot)$ is proposed to generate 2D feature planes for our detection-based formulation. Specifically, given $\Voxel^{(l)}_{\text{enc}}$, we learn per-voxel dense weight with a linear layer $\bm{\eta}=\mathbf{W}_{{d}}(\Voxel^{(l)}_{\text{enc}}), \bm{\eta} \in \mathbb{R}^{ X\times Y \times Z \times 3}$, which guides the adaptive pooling for the three TPV planes. Then, we conduct pooling across each axis ($X,Y,Z$) to generate projected 2D feature planes:
\begin{equation}
\label{eq:proj}\Pool_{{d}}^Z,\Pool_{{d}}^Y,\Pool_{{d}}^X(\Voxel)
=\sum_{z=1}^Z\mathcal{S}_{Z}(\bm{\eta}_{[:,:,:,1]})\circ\Voxel_{[:,:,z]},
\sum_{y=1}^Y\mathcal{S}_{Y}(\bm{\eta}_{[:,:,:,2]})\circ\Voxel_{[:,y,:]},
\sum_{x=1}^X\mathcal{S}_{X}(\bm{\eta}_{[:,:,:,3]})\circ\Voxel_{[x,:,:]}.
\end{equation}
Here, $\Pool_{{d}}^{(A)}$ indicates the $(A)$-axis projection, $\mathcal{S}_{(A)}$ is $\operatorname{softmax}$ on the $(A)$ axis, and $\circ$ is Hadamard product. Next, we deploy a 2D refinement layer to generate TPV feature at $XY, XZ, YZ$ planes:
\begin{equation}
\begin{split}
 {\Voxel}^{XY}_{\text{cls}}& = \Conv^{XY}_{\text{cls}}(\Pool_{d}^Z(\Voxel^{(l)}_{\text{enc}})), {\Voxel}^{XZ}_{\text{cls}} = \Conv^{XZ}_{\text{cls}}(\Pool_{d}^Y(\Voxel^{(l)}_{\text{enc}})),{\Voxel}^{YZ}_{\text{cls}} = \Conv^{YZ}_{\text{cls}}(\Pool_{d}^X(\Voxel^{(l)}_{\text{enc}})),\\
{\Voxel}^{XY}_{\text{reg}} &= \Conv^{XY}_{\text{reg}}(\Pool_{d}^{Z}(\Voxel^{(l)}_{\text{enc}})), {\Voxel}^{XZ}_{\text{reg}} = \Conv^{XZ}_{\text{reg}}(\Pool_{d}^Y(\Voxel^{(l)}_{\text{enc}})),{\Voxel}^{YZ}_{\text{reg}} = \Conv^{YZ}_{\text{reg}}(\Pool_{\text{ada}}^X(\Voxel^{(l)}_{\text{enc}})).
\end{split}
\end{equation}
Here, $\Conv^{(P)}_{(T)}(\cdot)$ is 2D convolution module for plane $(P)$ and task $(T)$. Compared with the conventional TPV pooling~\cite{TPVFormer}, our dense projection can adaptively discover the essential voxels during the 3D-to-2D dimensional reduction, thereby mitigating the information loss for the dense perception.

\noindent\textbf{Spatial Decoupling.} With projected 2D features, we spatially decouple it in each TPV plane with different spatial offsets, encouraging two tasks to focus on task-specific regions. Then, we use a $\Conv$ layer to fuse the decoupled features with the same expanded size, which is carried out as follows:
\begin{equation}
\label{eq:decouple}\begin{split}
 \Tilde{\Voxel}^{(l)}_{\text{cls}} &= \Conv^{\text{fuse}}_{\text{cls}}\left(\Defconv^{XY}_{\text{cls}}(\Voxel^{XY}_{\text{cls}})+ \Defconv^{XZ}_{\text{cls}}(\Voxel^{XZ}_{\text{cls}})+ \Defconv^{YZ}_{\text{cls}}(\Voxel^{YZ}_{\text{cls}})\right);\\
\Tilde{\Voxel}^{(l)}_{\text{reg}} &= \Conv^{\text{fuse}}_{\text{reg}}\left(\Defconv^{XY}_{\text{reg}}(\Voxel^{XY}_{\text{reg}})+ \Defconv^{XZ}_{\text{reg}}(\Voxel^{XZ}_{\text{reg}})+ \Defconv^{YZ}_{\text{reg}}(\Voxel^{YZ}_{\text{reg}})\right).
 \end{split}
\end{equation}
Here, \(\Defconv(\cdot)\) is the 2D deformable convolution module, which spatially decouples the dense voxel features for the two tasks. This decoupling reduces the computational burden and preserves task-driven spatial context, effectively addressing the misaligned feature preference of the two tasks. Finally, we send the decoupled multi-level features to the lightweight FPN~\cite{FPN} branches, respectively, to improve the task-specific learning, and collect the last-layer output with the same resolution as $\Voxel$ for the two tasks, which is denoted as: $\Voxel_{\text{cls}}= \text{FPN}(\{\Tilde{\Voxel}^{(l)}_{\text{cls}}\}_{l=1}^{L})$ and $\Voxel_{\text{reg}}= \text{FPN}(\{\Tilde{\Voxel}^{(l)}_{\text{reg}}\}_{l=1}^{L})$.

\subsection{Task-decoupled Dense Predictor}
Then, the decoupled features $\Voxel_\text{cls}$ and $\Voxel_\text{reg}$ are sent to the classification and regression branches (see Fig.~\ref{fig:overview}c). Here, we first densely regress the instance boundary to identify 3D objects, then aggregate instance-level semantics in the classification branch, achieving an instance-centric prediction.

\begin{figure*}[t]
 \centering
 \includegraphics[width=0.98\linewidth]{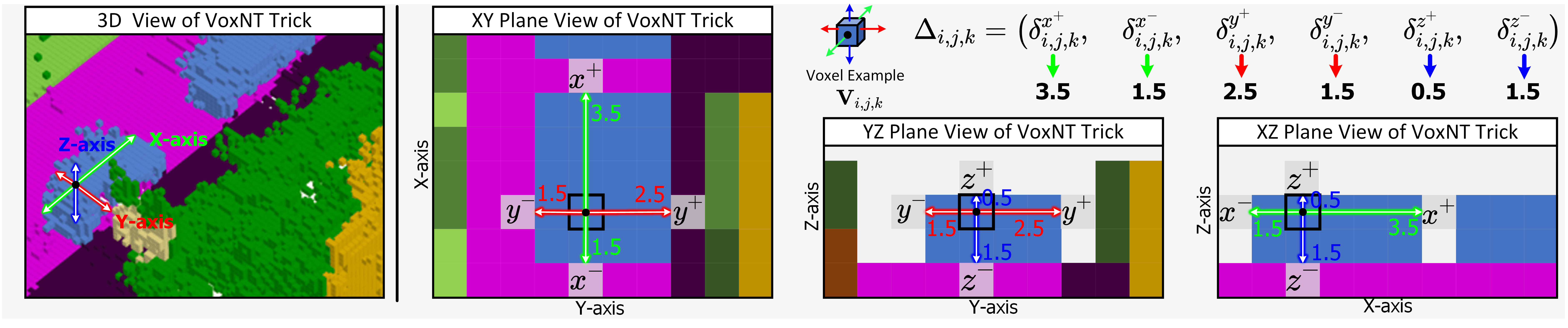}\vspace{-8pt}
 \caption{Illustration of our regression objective. For each voxel $\Voxel_{i,j,k}$, we scan in 6 directions and calculate the distance to the instance boundary to generate labels $\hat{\Delta}_{i,j,k}$, using the free-lunch (Sec.~\ref{sec:motivation}).}
 \label{fig:trick}
\vspace{-11pt}
\end{figure*}

\noindent\textbf{Regression Branch.} Given the volume for regression $\Voxel_\text{reg} \in \mathbb{R}^{X\times Y\times Z\times C }$, we aim to regress the distance to the instance boundary for each voxel $\Voxel_{i,j,k} \in \Voxel_\text{reg}$ (see Fig.~\ref{fig:trick}) in six directions, which form a 4D offset field $\Delta \in \mathbb{R}^{X \times Y \times Z \times 6}$. We use $6$ directions as it is the minimum number to determine a 3D bounding box, similar to the 2D bounding boxes determined with 4 directions~\cite{tian2019fcos}. Thus, considering the voxel in $\Voxel_\text{reg}$ at position $(i,j,k)$, the spatially associated element $\Delta_{i,j,k}$ in the offset field represents a 6-channel vector for the specific offset $\delta$ along the $X$, $Y$, and $Z$ axes, which is divided into positive ($^+$) and negative ($^-$) directions. This can be denoted as follows,
\begin{equation}
\Delta = \text{RegressionHead}(\Voxel_{\text{reg}}), \, \text{where} \, \Delta_{i,j,k}
= \bigl(
\delta_{i,j,k}^{x^+},
\delta_{i,j,k}^{x^-},
\delta_{i,j,k}^{y^+},
\delta_{i,j,k}^{y^-},
\delta_{i,j,k}^{z^+},
\delta_{i,j,k}^{z^-}
\bigr)
\in\mathbb{R}^6.
\end{equation}
Here, $\text{RegressionHead}(\cdot)$ is a lightweight head to predict the offset field, which is simply deployed as a two-convolution module to enable non-linearity with 6-channel output, followed by $\texttt{Sigmoid}$ for normalization. Compared with the anchor-based design~\cite{ren2016faster}, this design is highly computationally efficient due to its fully convolutional nature when deployed in the dense 3D voxel space. 

\noindent\textbf{Regression Loss.} In this task, we only have dense class labels for the voxel grid \({Y} \in \mathbb{N}^{X\times Y\times Z}\) without instance-level labels for regression, such as bounding boxes. Hence, to break through this, we propose a \textbf{Voxel-to-Instance (VoxNT) Trick} to freely transform the class labels to the instance-level offsets, fully using the free lunch in Sec.~\ref{sec:motivation}. The algorithm details are in the Appendix.

Fig.~\ref{fig:trick} shows the \Trick trick. In brief, for each voxel $\Delta_{i,j,k}$, e.g., the blue example, we scan across 6 directions \(\mathbf{d} \in \{x^+, x^-, y^+, y^-, z^+, z^-\}\) in labels ${Y}$, and stop when the class of next scanned voxel changes, indicating approaching the border. Then, we save the scanning distance as the offset labels: $\hat{\Delta}\in\mathbb{R}^{X\times Y\times Z \times 6}$. For stability, we round up $\hat{\Delta}$ to the integer and normalize into $[0,1]$ via the volume size. Finally, we deploy L1 loss to optimize the regression with enough gradient:
\begin{equation}
\label{eq:reg_loss}\mathcal{L}_{\text{reg}}
=\Sigma_{i,j,k,d}^{X,Y,Z,6}|\Delta_{i,j,k,d}-\hat\Delta_{i,j,k,d}|,
\end{equation}
where $\Delta \in \mathbb{R}^{X\times Y\times Z\times 6}$ is the predicted 4D offset field. Surprisingly, as a by-product, the offset field can identify wrongly labeled dynamic objects (like cars) for more accurate training (see Appendix).

\noindent\textbf{Classification Branch.} Based on the regressed offset $\Delta$, we aim to enhance instance-level perception by aggregating the semantics within instances. A natural idea~\cite{girshick2015fast} is to crop 3D bounding boxes, which is extremely computationally expensive. Hence, we propose an alternative solution by adaptively aggregating the voxels at regressed positions, considering the informative nature of borders~\cite{law2018cornernet,zhou2019bottom}. For each voxel $\Voxel_{i,j,k}\in\Voxel_{\text{cls}}$, we have the predicted offsets $\Delta_{i,j,k}$ associating with 6 voxels, which are extracted as a instance-level voxel set $\Voxel_{\delta}$, where $\delta \in \Delta_{i,j,k}$ and $|\Delta_{i,j,k}|$=6. Then, each voxel $\Voxel_{i,j,k}$ will query the semantics from the associated voxel set $\Voxel_{\delta}$ attentively:
\begin{equation}
 \label{eq:adaptive} \hat{\Voxel}_{i,j,k}=
\text{Norm}(\sum_{\delta\in\Delta_{i,j,k}}
\frac{\exp\bigl(\mathbf{W}_q\Voxel_{i,j,k})^\top (\mathbf{W}_k\Voxel_\delta/\sqrt{d_k}\bigr)}
{\sum_{\delta'\in\Delta_{i,j,k}}\exp\bigl(\mathbf{W}_q\Voxel_{i,j,k})^\top (\mathbf{W}_k\Voxel_{\delta'}/\sqrt{d_k}\bigr)}
\,\mathbf{W}_v\Voxel_\delta + \Voxel_{i,j,k}).
\end{equation}
Here, $\hat{\Voxel}_{i,j,k} \in \hat{\Voxel}_{\text{cls}}$ is refined feature, each $\mathbf{W}_{(\cdot)}$ is a linear layer, $d_k$ is the channel number, and $\text{Norm}$ is $\texttt{Group Normalization}$. With $N=4$ aggregation layers, each voxel is able to gain sufficient perception of the whole instance, enhancing instance-level semantics. This design is justified in Fig.~\ref{fig:agg} and the Appendix. Finally, $\hat{\Voxel}_{\text{cls}}$ is sent to a simple classification head to generate class predictions.


\noindent\textbf{Classification Loss.} In dense detection~\cite{tian2019fcos}, classification is optimized as segmentation in a per-pixel dense manner. Hence, to optimize the classification branch in our \Method, we naturally deploy the dense semantic prediction loss following the previous arts~\cite{MonoOcc,yucontext,OccFormer}:
\begin{equation}
\label{eq:cls_loss}\mathcal{L}_{\text{cls}} = \mathcal{L}_{\text{ce}} + \mathcal{L}_{\text{aff}}^{\text{bin}} + \mathcal{L}_{\text{aff}}^{\text{sem}},
\end{equation}
where $\mathcal{L}_{\text{ce}}$ is the cross-entropy loss weighted by class frequencies, and $\mathcal{L}_{\text{aff}}^{\text{bin}}$ and $ \mathcal{L}_{\text{aff}}^{\text{sem}}$ are affinity loss with binary and semantic settings. We set the consistent loss weight as 1.0 for convenience.

\subsection{Optimization}
To train the proposed \Method, we implement the whole optimization objective written as follows,
\begin{equation}
 \mathcal{L}_{\text{\Method}} = \mathcal{L}_{\text{cls}} + \mathcal{L}_{\text{reg}} + \lambda \mathcal{L}_{\text{occ}}^{\text{aux}},
\end{equation}
where $ \mathcal{L}_{\text{cls}}$ is dense classification loss (Eq.~\ref{eq:cls_loss}) and $ \mathcal{L}_{\text{reg}}$ is dense regression loss (Eq.~\ref{eq:reg_loss}). Following the baseline~\cite{MonoOcc}, we retain the voxel-centric segmentation loss (cross‐entropy and affinity terms) applied to $\Voxel$ as an auxiliary prior constraint ($\mathcal{L}_{\text{occ}}^{\text{aux}}$) before instance-centric learning, which can stabilize the optimization. This also enhances the multiple-run robustness, which is justified in the Appendix. The weighting factor $\lambda=0.2$ is empirically adjusted to balance the model learning at the instance level.

\section{Experiments}

\begin{table*}
\newcommand{\clsname}[2]{
\rotatebox{90}{
\hspace{-6pt}
\textcolor{#2}{$\blacksquare$}
\hspace{-6pt}
\renewcommand\arraystretch{0.6}
\begin{tabular}{l}
#1                                      \\
\hspace{-4pt} ~\tiny(\semkitfreq{#2}\%) \\
\end{tabular}
}}
\centering\huge
\caption{Camera-based results on SemanticKITTI~\cite{SemanticKITTI} hidden test set. \textbf{T} is using extra temporal information. The best and the second best results are in \textbf{bold} and \underline{underlined}, respectively. R-50 indicates ResNet-50, and Eff-B7 is the stronger EfficientNet-B7 backbone with more parameters. $\dag$ indicates setting the same weight (3.0) on the cross-entropy loss as the recent work~\cite{wang2025l2cocc}. }
\resizebox{\linewidth}{!}
{
\begin{tabular}{l|c|c|cc|ccccccccccccccccccc}
\toprule		
Method 								 &
\textbf{Arch.}                         & 
\textbf{T}                           & 
IoU 								 & 
mIoU  								 &  
\clsname{road}{road}                 & 
\clsname{sidewalk}{sidewalk}         &        
\clsname{parking}{parking}           & 
\clsname{other-grnd.}{otherground}   & 
\clsname{building}{building}         & 
\clsname{car}{car} 					 & 
\clsname{truck}{truck}               &
\clsname{bicycle}{bicycle}           &
\clsname{motorcycle}{motorcycle}     &
\clsname{other-veh.}{othervehicle}   &
\clsname{vegetation}{vegetation}     &
\clsname{trunk}{trunk}               &
\clsname{terrain}{terrain}           &
\clsname{person}{person}             &
\clsname{bicyclist}{bicyclist}       &
\clsname{motorcyclist}{motorcyclist} &
\clsname{fence}{fence}               &
\clsname{pole}{pole}                 &
\clsname{traf.-sign}{trafficsign}
\\
\midrule
MonoScene$^\ast$~\cite{MonoScene}& Eff-B7& & 34.16 & 11.08 & 54.70 & 27.10 & 24.80 & 5.70 & 14.40 & 18.80 & 3.30 & 0.50 & 0.70 & 4.40  & 14.90 & 2.40  & 19.50 & 1.00  & 1.40 & 0.40  & 11.10 & 3.30 & 2.10         \\
TPVFormer~\cite{TPVFormer}  & Eff-B7  &    &34.25 & 11.26 & 55.10 & 27.20 & 27.40 & 6.50 & 14.80 & 19.20 & 3.70 & 1.00 & 0.50 & 2.30  & 13.90 & 2.60  & 20.40 & 1.10  & 2.40 & 0.30  & 11.00 & 2.90 & 1.50 		  \\
SurroundOcc~\cite{surroundOcc} &Eff-B7&   & 34.72 & 11.86 & 56.90 & 28.30 & 30.20 & 6.80 & 15.20 & 20.60 & 1.40 & 1.60 & 1.20 & 4.40  & 14.90 & 3.40  & 19.30 & 1.40  & 2.00 & 0.10  & 11.30 & 3.90 & 2.40         \\
OccFormer~\cite{OccFormer} &Eff-B7    &   & 34.53 & 12.32 & 55.90 & 30.30 & {31.50} & 6.50 & 15.70 & 21.60 & 1.20 & 1.50 & 1.70 & 3.20  & 16.80 & 3.90  & 21.30 & 2.20  & 1.10 & 0.20  & 11.90 & 3.80 & 3.70         \\
IAMSSC~\cite{IAMSSC}&R-50  & & 43.74 & 12.37 & 54.00 & 25.50 & 24.70 & 6.90 & 19.20 & 21.30 & 3.80 & 1.10 & 0.60 & 3.90 & 22.70 & 5.80 & 19.40 & 1.50 & 2.90 & 0.50 & 11.90 & 5.30 & 4.10 \\
VoxFormer~\cite{VoxFormer} &R-50  &    & 42.95 & 12.20 & 53.90 & 25.30 & 21.10 & 5.60
& 19.80 & 20.80 & 3.50 & 1.00 & 0.70 & 3.70  & 22.40 & 7.50  & 21.30 & 1.40  & 2.60 
& 0.20  & 11.10 & 5.10 & 4.90         \\
VoxFormer~\cite{VoxFormer}&R-50  &\checkmark      & 43.21 & 13.41 & 54.10 & 26.90 & 25.10 & 7.30 & 23.50 & 21.70 & 3.60 & 1.90 & 1.60 & 4.10 & 24.40 & 8.10 & 24.20 & 1.60 & 1.10 & 0.00 & 13.10 & 6.60 & 5.70 \\
DepthSSC~\cite{DepthSSC}&R-50   &        & {44.58} & 13.11 & 55.64 & 27.25 & 25.72 & 5.78
& 20.46 & 21.94 & 3.74 & 1.35 & 0.98 & 4.17  & 23.37 & 7.64  & 21.56 & 1.34  & 2.79
& 0.28  & 12.94 & 5.87 & 6.23         \\
Symphonize~\cite{Symphonize}& R-50 	&  & 42.19 & 15.04 & 58.40 & 29.30 & 26.90 & {11.70}
& {24.70} & 23.60 & 3.20 & 3.60 & {2.60} & {5.60}  & 24.20 & 10.00 & 23.10 & \Best{3.20}  & 1.90  
& \Best{2.00}  & 16.10 & {7.70} & 8.00         \\
HASSC~\cite{HASSC}&R-50 & & 43.40 & 13.34 & 54.60 & 27.70 & 23.80 & 6.20 & 21.10 & 22.80 & 4.70 & 1.60 & 1.00 & 3.90 & 23.80 & 8.50 & 23.30 & 1.60 & \SecondBest{4.00} & 0.30 & 13.10 & 5.80 & 5.50 \\
HASSC~\cite{HASSC}&R-50  &\checkmark & 42.87 & 14.38 & 55.30 & 29.60 & 25.90 & 11.30 & 23.10 & 23.00 & 2.90 & 1.90 & 1.50 & 4.90 & 24.80 & 9.80 & 26.50 & 1.40 & 3.00 & 0.00 & 14.30 & 7.00 & 7.10 \\
StereoScene~\cite{StereoScene} &Eff-B7 & & 43.34 & 15.36 & {61.90} & {31.20} & 30.70 & 10.70 & 24.20 & 22.80 & 2.80 &  3.40 & {2.40} & \Best{6.10} & 23.80 & 8.40 & {27.00} & {2.90} & 2.20 & 0.50 & 16.50 & 7.00 & 7.20 \\
H2GFormer~\cite{H2GFormer}&R-50 & & 44.20 & 13.72 & 56.40 & 28.60 & 26.50 & 4.90 & 22.80 & 23.40 & 4.80 & 0.80 & 0.90 & 4.10 & 24.60 & 9.10 & 23.80 & 1.20 & 2.50 & 0.10 & 13.30 & 6.40 & 6.30 \\
H2GFormer~\cite{H2GFormer}&R-50 & \checkmark & 43.52 & 14.60 & 57.90 & 30.40 & 30.00 & 6.90 & 24.00 & 23.70 & {5.20} & 0.60 & 1.20 & 5.00 & {25.20} & {10.70} & 25.80 & 1.10 & 0.10 & 0.00 & 14.60 & 7.50 & {9.30} \\
MonoOcc~\cite{MonoOcc}&R-50  & & - & 13.80 & 55.20 & 27.80 & 25.10 & 9.70 & 21.40 & 23.20 & {5.20} & 2.20 & 1.50 & 5.40 & 24.00 & 8.70 & 23.00 & 1.70 & 2.00 & 0.20 & 13.40 & 5.80 & 6.40 \\
CGFormer~\cite{yucontext}&Eff-B7 & & 44.41 & {16.63} & {64.30} & {34.20} & {34.10} & {12.10} & {25.80} & {26.10} & 4.30 & {3.70} & 1.30 & 2.70 & 24.50 & {11.20} & {29.30} & 1.70 & {3.60} & 0.40 & {18.70} & {8.70} & {9.30} \\
L2COcc-C~\cite{wang2025l2cocc}&Eff-B7 & &44.31&17.03&\Best{66.00}&35.00&33.10&\SecondBest{13.50}&25.10&27.20&3.00&3.50&3.60&4.30&25.20&11.50&30.10&1.50&2.40&0.20&20.50&{9.10}&8.90 \\
HTCL~\cite{li2024hierarchical} &Eff-B7 &  \checkmark&44.23 & {17.09} & {64.40} & 34.80 & 33.80 & {12.40} & {25.90} & \Best{27.30} & \SecondBest{5.70} & 1.80 & 2.20 & 5.40 & {25.30} & 10.80 & {31.20} & 1.10 & 3.10 & 0.90 & {21.10} & {9.00} & 8.30 \\
\midrule
\textbf{\Method (Ours)} &R-50& & \SecondBest{47.27} & \SecondBest{18.47} & {64.70} & \SecondBest{35.50} & \SecondBest{34.80} & \Best{14.40} & \SecondBest{28.10} & \SecondBest{26.90} & \Best{6.10} & \Best{5.90} & \SecondBest{5.10} & {5.00} & \SecondBest{28.70} & \Best{13.60} & \SecondBest{31.70} & \SecondBest{3.10} & \SecondBest{4.00} & \SecondBest{1.30} & \SecondBest{21.50} & \SecondBest{10.10} & \SecondBest{10.30} \\
\textbf{$\text{VoxDet}^\dag$ (Ours)}&R-50& &\Best{47.81}& \Best{18.67}  & \SecondBest{65.50} & \Best{36.10} & \Best{35.50} & 13.20 & \Best{28.40} & \Best{27.30} & 5.40 & \SecondBest{4.60} & \Best{5.40} & \SecondBest{5.40} & \Best{29.50} & \SecondBest{13.10} & \Best{32.00} & \SecondBest{3.10} & \Best{6.10} & 0.90 & \Best{22.10} & \Best{10.20} & \Best{11.10}\\
\bottomrule
\end{tabular} 
}
\setlength{\abovecaptionskip}{0cm}
\setlength{\belowcaptionskip}{0cm}
\label{tab:sem_kitti_test}
\vspace{-4mm}
\end{table*}

\begin{table}
\newcommand{\clsname}[2]{
\rotatebox{90}{
\hspace{-6pt}
\textcolor{#2}{$\blacksquare$}
\hspace{-6pt}
\renewcommand\arraystretch{0.6}
\begin{tabular}{l}
#1                                       \\
\hspace{-4pt} ~\tiny(\sscbkitfreq{#2}\%) \\
\end{tabular}
}}
\centering\huge
\caption{Camera-based results on SSCBench-KITTI360 test set. $^*$ is corrected using the consistent dataset version~\cite{guo2025sgformer}. The best and second best results are are in \Best{bold} and \SecondBest{underlined}, respectively.}\vspace{+3pt}
\resizebox{\linewidth}{!}
{
\begin{tabular}{l|cc|cccccccccccccccccc}
\toprule
Method                            			&
IoU                                 		&
mIoU                                		&
\clsname{car}{car}                      	&
\clsname{bicycle}{bicycle}              	&
\clsname{motorcycle}{motorcycle}        	&
\clsname{truck}{truck}                   	&
\clsname{other-veh.}{othervehicle}      	&
\clsname{person}{person}                	&
\clsname{road}{road}                    	&
\clsname{parking}{parking}              	&
\clsname{sidewalk}{sidewalk}            	&
\clsname{other-grnd.}{otherground}      	&
\clsname{building}{building}            	&
\clsname{fence}{fence}                  	&
\clsname{vegetation}{vegetation}        	&
\clsname{terrain}{terrain}              	&
\clsname{pole}{pole}                    	&
\clsname{traf.-sign}{trafficsign}       	&
\clsname{other-struct.}{otherstructure} 	&
\clsname{other-obj.}{otherobject}
\\
\midrule
\multicolumn{21}{l}{\textit{LiDAR-based methods}}                                                                                                        \\
\hline
\textcolor{gray}{SSCNet~\cite{SSCNet}}        & \textcolor{gray}{{53.58}} & \textcolor{gray}{16.95} & \textcolor{gray}{{31.95}} & \textcolor{gray}{0.00} & \textcolor{gray}{0.17} & \textcolor{gray}{10.29} & \textcolor{gray}{0.00} & \textcolor{gray}{0.07} & \textcolor{gray}{{65.70}}
& \textcolor{gray}{{17.33}} & \textcolor{gray}{{41.24}} & \textcolor{gray}{3.22} & \textcolor{gray}{{44.41}} & \textcolor{gray}{6.77} & \textcolor{gray}{{43.72}} & \textcolor{gray}{{28.87}} & \textcolor{gray}{0.78}  & \textcolor{gray}{0.75} & \textcolor{gray}{8.69} & \textcolor{gray}{0.67} \\
\textcolor{gray}{LMSCNet~\cite{LMSCNet}}      & \textcolor{gray}{47.35} & \textcolor{gray}{13.65} & \textcolor{gray}{20.91} & \textcolor{gray}{0.00} & \textcolor{gray}{0.00} & \textcolor{gray}{0.26}  & \textcolor{gray}{0.58} & \textcolor{gray}{0.00} & \textcolor{gray}{62.95}        
& \textcolor{gray}{13.51} & \textcolor{gray}{33.51} & \textcolor{gray}{0.20} & \textcolor{gray}{43.67} & \textcolor{gray}{0.33} & \textcolor{gray}{40.01} & \textcolor{gray}{26.80} & \textcolor{gray}{0.00}  & \textcolor{gray}{0.00} & \textcolor{gray}{3.63} & \textcolor{gray}{0.00} \\
\specialrule{0.7pt}{0pt}{0pt}
\multicolumn{21}{l}{\textit{Camera-based methods}}     \\                                                            
\hline
MonoScene~\cite{MonoScene}  & 37.87 & 12.31 & 19.34 & 0.43 & 0.58 & 8.02  & 2.03 & 0.86 & 48.35        
& 11.38 & 28.13 & 3.32 & 32.89 & 3.53 & 26.15 & 16.75 & 6.92 & 5.67 & 4.20 & 3.09  \\
TPVFormer~\cite{TPVFormer}  & 40.22 & 13.64 & 21.56 & 1.09 & 1.37 & 8.06  & 2.57 & 2.38 & 52.99        
& 11.99 & 31.07 & 3.78 & 34.83 & 4.80 & 30.08 & 17.52 & 7.46 & 5.86 & 5.48 & 2.70  \\
OccFormer~\cite{OccFormer}  & 40.27 & 13.81 & 22.58 & 0.66 & 0.26 & 9.89  & 3.82 & 2.77 & 54.30       
& 13.44 & 31.53 & 3.55 & 36.42 & 4.80 & 31.00 & 19.51 & 7.77 & 8.51 & 6.95 & 4.60  \\
VoxFormer~\cite{VoxFormer}  & 38.76 & 11.91 & 17.84 & 1.16 & 0.89 & 4.56  & 2.06 & 1.63 & 47.01
& 9.67  & 27.21 & 2.89 & 31.18 & 4.97 & 28.99 & 14.69 & 6.51 & 6.92 & 3.79 & 2.43  \\
IAMSSC~\cite{IAMSSC}   & 41.80 & 12.97 & 18.53 & {2.45} & 1.76 & 5.12 & 3.92 & 3.09 & 47.55 & 
10.56 & 28.35 & 4.12 & 31.53 & 6.28 & 29.17 & 15.24 & 8.29 & 7.01 & 6.35 & 4.19 \\
DepthSSC~\cite{DepthSSC}    & 40.85 & 14.28 & 21.90 & 2.36 & {4.30} & 11.51 & 4.56 & 2.92 & 50.88
& 12.89 & 30.27 & 2.49 & {37.33} & 5.22 & 29.61 & {21.59} & 5.97 & 7.71 & 5.24 & 3.51  \\
Symphonies$^*$~\cite{Symphonize} & {43.41} & {17.82} & {26.86} & \SecondBest{4.21} & \SecondBest{4.90} & {14.20} & \SecondBest{7.76} & {6.57} & {57.30}
& {13.58} & {35.24} & {4.57} & 39.20 & {7.95} & {34.33} & 19.19 & {14.04} & {15.78} & {8.23} & {6.04} \\
SGN-S~\cite{SGN} &46.22& 17.71&  28.20 & 2.09 & 3.02 & 11.95 & 3.68 & 4.20 & 59.49 & 14.50 & 36.53 & 4.24 & 39.79 & 7.14 & 36.61 & 23.10 & 14.86 & 16.14 & 8.24 & 4.95 \\
SGN-T~\cite{SGN}  &47.06 &18.25& 29.03 & 3.43 & 2.90 & 10.89 & 5.20 & 2.99 & 58.14 & 15.04 & 36.40 & 4.43 & 42.02 & 7.72 & 38.17 & 23.22 & \SecondBest{16.73} & 16.38 & 9.93 & 5.86 \\ 
CGFormer~\cite{yucontext} & \SecondBest{48.07} & \SecondBest{20.05} & \SecondBest{29.85} & {3.42} & 3.96 & \SecondBest{17.59} & {6.79} & \SecondBest{6.63} & \Best{63.85} & \SecondBest{17.15} & \Best{40.72} & \SecondBest{5.53} & {42.73} & \SecondBest{8.22} & {38.80} & \SecondBest{24.94} & {16.24} & \SecondBest{17.45} & \SecondBest{10.18} & \SecondBest{6.77} \\
SGFormer~\cite{guo2025sgformer}&46.35 & 18.30& 27.80 & 0.91 & 2.55 & 10.73 & 5.67 & 4.28 & 61.04 & 13.21 & 37.00 & 5.07 & \SecondBest{43.05} & 7.46 & \SecondBest{38.98} & 24.87 & 15.75 & 16.90 & 8.85 & 5.33 \\
\midrule
 \textbf{\Method (Ours)} & \Best{{48.59}} & \Best{{21.40}}  & \Best{29.92} & \Best{5.13} & \Best{8.36} & \Best{19.13} & \Best{8.04} & \Best{7.84} & \SecondBest{62.83} & \Best{18.99} & \SecondBest{40.10} & \Best{5.58} & \Best{44.47} & \Best{10.62} & \Best{39.03} & \Best{26.16} & \Best{18.19} & \Best{20.78} & \Best{11.66} & \Best{8.34}\\
\bottomrule
\end{tabular}
}
\label{tab:kitti_360}
\vspace{-10pt}
\end{table}

\begin{table}[t]
\newcommand{\clsname}[2]{
\rotatebox{90}{
\hspace{-6pt}
\textcolor{#2}{$\blacksquare$}
\hspace{-6pt}
\renewcommand\arraystretch{0.6}
\begin{tabular}{l}
#1                                      \\
\hspace{-4pt} ~\tiny(\semkitfreq{#2}\%) \\
\end{tabular}
}}
\centering\huge
\caption{LiDAR-based results on SemanticKITTI~\cite{SemanticKITTI} hidden test set with single frame (no extra temporal information) for fair comparison. Our \Method only uses point cloud input. The best and the second best results are in \textbf{bold} and \underline{underlined}, respectively. Previous SoTA is~\cite{wang2024voxel}.}\vspace{+5pt}
\resizebox{\linewidth}{!}
{
\begin{tabular}{l |l|cc|ccccccccccccccccccc}
\toprule		
Method 								 & 
\textbf{T} & 
IoU 								 & 
mIoU  								 &  
\clsname{road}{road}                 & 
\clsname{sidewalk}{sidewalk}         &        
\clsname{parking}{parking}           & 
\clsname{other-grnd.}{otherground}   & 
\clsname{building}{building}         & 
\clsname{car}{car} 					 & 
\clsname{truck}{truck}               &
\clsname{bicycle}{bicycle}           &
\clsname{motorcycle}{motorcycle}     &
\clsname{other-veh.}{othervehicle}   &
\clsname{vegetation}{vegetation}     &
\clsname{trunk}{trunk}               &
\clsname{terrain}{terrain}           &
\clsname{person}{person}             &
\clsname{bicyclist}{bicyclist}       &
\clsname{motorcyclist}{motorcyclist} &
\clsname{fence}{fence}               &
\clsname{pole}{pole}                 &
\clsname{traf.-sign}{trafficsign}
\\
\midrule
SSCNet~\cite{SSCNet}   &      & 29.8 &  9.5 & 27.6 & 17.0 & 15.6 &  6.0 & 20.9 & 10.4 &  1.8 &  0.0 &  0.0 &  0.1 & 25.8 & 11.9 & 18.2 &  0.0 &  0.0 &  0.0 & 14.4 &  7.9 &  3.7 \\
SSCNet-full~\cite{SSCNet} &   & 50.0 & 16.1 & 51.2 & 30.8 & 27.1 &  6.4 & 34.5 & 24.3 &  1.2 &  0.5 &  0.8 &  4.3 & 35.3 & 18.2 & 29.0 &  0.3 &  0.3 &  0.0 & 19.9 & 13.1 &  6.7 \\
TS3D~\cite{garbade2019two} &  & 29.8 &  9.5 & 28.0 & 17.0 & 15.7 &  4.9 & 23.2 & 10.7 &  2.4 &  0.0 &  0.0 &  0.2 & 24.7 & 12.5 & 18.3 &  0.0 &  0.1 &  0.0 & 13.2 &  7.0 &  3.5 \\
TS3D/DNet~\cite{behley2019semantickitti} & & 25.0 & 10.2 & 27.5 & 18.5 & 18.9 &  6.6 & 22.1 &  8.0 &  2.2 &  0.1 &  0.0 &  4.0 & 19.5 & 12.9 & 20.2 &  2.3 &  0.6 &  0.0 & 15.8 &  7.6 &  7.0 \\
LMSCNet~\cite{LMSCNet}   &    & 55.3 & 17.0 & 64.0 & 33.1 & 24.9 &  3.2 & 38.7 & 29.5 &  2.5 &  0.0 &  0.0 &  0.1 & 40.5 & 19.0 & 30.8 &  0.0 &  0.0 &  0.0 & 20.5 & 15.7 &  0.5 \\
LMSCNet-SS~\cite{LMSCNet} &   & 56.7 & 17.6 & 64.8 & 34.7 & 29.0 &  4.6 & 38.1 & 30.9 &  1.5 &  0.0 &  0.0 &  0.8 & 41.3 & 19.9 & 32.1 &  0.0 &  0.0 &  0.0 & 20.5 & 15.7 &  0.8 \\
Local-DIFs~\cite{rist2021semantic} &  & 57.7 & 22.7 & 67.9 & {42.9} &\Best{40.1} & 11.4 & 40.4 & 34.8 &  4.4 &  3.6 &  2.4 &  4.8 & 42.2 & 26.5 & 39.1 &  2.5 &  1.1 &  0.0 & 29.0 & 21.3 & 17.5 \\
JS3C-Net~\cite{yan2021sparse} &     & 56.6 & {23.8} & 64.7 & 39.9 & 34.9 & \SecondBest{14.1} & 39.4 & 33.3 & \Best{7.2} & \Best{14.4} &  \SecondBest{8.8} & \Best{12.7} & 43.1 & 19.6 & {40.5} &  \Best{8.0} &  \Best{5.1} &  0.4 & 30.4 & 18.9 & 15.9 \\
SSA-SC~\cite{yang2021semantic}  &     & 58.8 & 23.5 & {72.2} & \SecondBest{43.7} & {37.4} & 10.9 & {43.6} & 36.5 &  5.7 & \SecondBest{13.9} &  4.6 &  7.4 & {43.5} & 25.6 & \SecondBest{41.8} &  \SecondBest{4.4} &  2.6 &  0.7 & {30.7} & 14.5 &  6.9 \\
L2COcc-D~\cite{wang2025l2cocc} &      & 45.3 & 18.1 & 68.2 & 36.9 & 34.6 & \Best{16.2} & 25.8 & 28.3 &  4.5 &  4.9 &  3.3 &  7.2 & 26.2 & 11.9 & 32.0 &  2.1 &  2.4 &  0.3 & 21.6 &  9.6 &  9.5 \\
L2COcc-L~\cite{wang2025l2cocc}  &     & {60.3} & 23.3 & {68.5} & 40.6 & 33.2 &  6.1 & 41.5 & \SecondBest{36.8} &  5.4 &  8.7 &  4.1 &  9.0 & 42.6 &{28.7} & 36.9 &  1.4 &  2.9 &  {1.0} & {27.7} & \SecondBest{27.0} &\SecondBest{21.9} \\
OccMamba~\cite{li2024occmamba} & \checkmark  &  -& 24.6& - & -  & - &  -  &-  & -  & -  &  - &  -  & - &-  &-  & - & - &  -  &-  & -  & -  &-  \\
VPNet~\cite{wang2024voxel} & \checkmark& \SecondBest{60.4} & \SecondBest{25.0} & \SecondBest{72.4} & \Best{44.3} & 40.5 & 14.8 & \SecondBest{44.0} & 37.2 & 4.3 & 14.0 & \Best{9.8} & 8.2 & \SecondBest{45.3} & \Best{30.9} & 42.1 & 4.9 & 2.0 & \Best{2.4} & \SecondBest{32.7} & 17.1 & 8.8 \\
\midrule
\Best{VoxDet-L (Ours)} & &\Best{63.0} & \Best{26.0} & \Best{73.0} & {43.6} & \SecondBest{37.5} & 10.3 & \Best{44.5} & \Best{37.7} & \SecondBest{6.6} & 9.9 & 6.2 & \SecondBest{11.8} &\Best{45.9} & \SecondBest{30.7} & \Best{43.5} & 2.7 & \SecondBest{3.2} & \SecondBest{1.3} & \Best{34.0} & \Best{27.8} & \Best{23.7} \\
\bottomrule
\end{tabular} 
}
\setlength{\abovecaptionskip}{0cm}
\setlength{\belowcaptionskip}{0cm}
\label{tab:lidar-sem_kitti_test}
\end{table}

\subsection{Comparison with State-of-the-art Methods} 

\subsubsection{Camera-based Benchmarks}

Tab.~\ref{tab:sem_kitti_test} shows the comparison on the hidden test set of SemanticKITTI~\cite{SemanticKITTI}. \Method achieves new records with an IoU of 47.27 and mIoU of 18.47. Compared with the methods using additional temporal labels like VoxFormer-T~\cite{VoxFormer}, HASSC-T~\cite{HASSC}, H2GFormer-T~\cite{H2GFormer}, and HTCL-T~\cite{li2024hierarchical}, \Method comprehensively surpasses them and gives noticeable 3.04 and 1.38 gains on IoU and mIoU over the previous best entry~\cite{li2024hierarchical}. This clearly verifies the effectiveness of our method.

We further list the comparison on SSCBench-KITTI-360~\cite{KITTI360} in Tab.~\ref{tab:kitti_360}. \Method achieves an IoU of 48.59 and mIoU of 21.40, outperforming other camera-based methods. \emph{Notably, VoxDet achieves the best on all instance-related classes, showing an impressive capacity in understanding outdoor agents.}

\subsubsection{Extension to LiDAR-based Benchmark}

In Tab.~\ref{tab:lidar-sem_kitti_test}, we deploy our method with LiDAR settings, denoted as \textbf{\Method-L}, and give a comparison on the hidden test set (online evaluation). Our VoxDet-L achieves the new record with a 63.0 IoU and 26.0 mIoU, significantly surpassing state-of-the-art methods~\cite{wang2025l2cocc,wang2024voxel}. Our method achieves the best on large-scale objects like \emph{building, vegetation}, middle-scale objects like \emph{car}, and small-scale objects such as \emph{pole, traffic-sign}. \highlight{Additionally, our method is also state-of-the-art in terms of model efficiency.} Our VoxDet-L (with only 22.1 M parameters) significantly surpasses the latest counterpart VPNet (with 35.8 M parameters)~\cite{wang2024voxel} published on NeurIPS 2024 and L2COcc-L (with 36.2 M parameters) published on ArXiv 25 on IoU, mIoU, and model parameters. 

{Notably, with our effective regression-include formulation, \Method-L achieves a significantly higher IoU of 63.0 (the completion metric) without using extra labels/data/temporal information/models, ranking 1st on the CodaLab leaderboard\footnote{Online leaderboard: https://codalab.lisn.upsaclay.fr/competitions/7170\#results} (Updated on May 22).} This capability is essentially crucial for autonomous navigation in preventing collisions with obstacles.

\subsection{Quantitative Study}

\begin{table*}[t]
\small
\centering
\begin{minipage}[t]{0.46\linewidth}
\centering
\caption{Ablation study on each key module.}\vspace{-3pt}
\resizebox{1\textwidth}{!}{
\begin{tabular}{c|cc|cc|cc |c}
\toprule
&\multicolumn{2}{c|}{TDP}& \multicolumn{2}{c|}{SVE}& IoU & mIoU  &  N$_{\mathrm{param}}$ \\
& REG & CLS &  REG & CLS & (\%)&(\%) & (M)\\
\midrule
(a) & & & & &42.71 & 16.28& 48.7 \\
(b) & \cmark& & &  &45.42 & 16.42& 48.9  \\
(c) & \cmark& \cmark& & & 46.79 & 17.85 & 49.4\\
(d) & \cmark& \cmark& \cmark& &  47.14& 18.02 &51.1\\
(e) & \cmark& \cmark& & \cmark& 47.08 &18.27 &51.1\\
(f) & \cmark& \cmark& \cmark& \cmark& \Best{47.36}  & \Best{18.73}& 52.8 \\
\bottomrule
\end{tabular}}
\label{tab:ablation}
\end{minipage}
\hfill
\begin{minipage}[t]{0.46\linewidth}
\centering
\caption{Further analysis of varied designs.}\vspace{-3pt}
\resizebox{1\textwidth}{!}{
\begin{tabular}{c|l|cc}
\toprule
& Detailed designs & IoU (\%)& mIoU (\%) \\ \midrule
\multirow{2}{*}{{TDP}}&   $\Delta$ guidance $\rightarrow$ Self-attention & 46.82 & 18.15   \\
& Eq.~\ref{eq:adaptive} $\rightarrow$ Weighted fusion & 47.01  &18.38 \\
\hline
\multirow{4}{*}{{SVE}}
& Eq.~\ref{eq:proj}$\rightarrow$ Average pooling &  47.18 & 18.49\\
& $\text{DefConv}$ $\rightarrow$ \text{Conv} & 46.92 & 18.08\\
& TPV $\rightarrow$ ResBlock  & 47.02  &  18.32 \\
& Task-decoupled $\rightarrow$ Task-shared  &  46.81 &  17.88 \\
\hline
&Full model& \Best{47.36}  & \Best{18.73} \\ 
\bottomrule
\end{tabular}}
\label{tab:further}
\end{minipage}
\vspace{-1mm}
\end{table*}

\begin{table*}[t]
\small
\centering
\begin{minipage}[t]{0.56\linewidth}
\centering
\caption{Efficiency comparison with SoTA methods.}\vspace{-2pt}
\resizebox{1.0\textwidth}{!}{
\begin{tabular}{l|cc|cc}
\toprule
Method & N$_{\mathrm{param}}$ $\downarrow$ & T$_{\mathrm{inf}}$ $\downarrow$& IoU (\%) $\uparrow$ & mIoU (\%) $\uparrow$ \\
\midrule
OccFormer~\cite{OccFormer}$_{[\mathrm{ICCV'23}]}$ & 214 & 199 & 36.42& 13.50 \\
StereoScene~\cite{StereoScene}$_{[\mathrm{IJCAI'24}]}$ & 117 & 258 & 43.85 & 15.43 \\
CGFormer~\cite{yucontext}$_{[\mathrm{NeurIPS'24}]}$ & 122 & 205 & 45.99 & 16.89 \\
SGFormer~\cite{guo2025sgformer}$_{[\mathrm{CVPR'25}]}$ & 126 & - & 45.01 & 16.68 \\
ScanSSC~\cite{bae2024three}$_{[\mathrm{CVPR'25}]}$ & 145 & 261 & 45.95 & 17.12 \\
\hline
\textbf{\Method (Ours)} & \Best{53} & \Best{159} & \Best{47.36} & \Best{18.73} \\		
\bottomrule
\end{tabular}
}
\label{tab:efficiency}
\end{minipage}
\hfill
\begin{minipage}[t]{0.4\linewidth}
\centering
\caption{Results with monocular depth.}\vspace{-2pt}
\resizebox{0.8\textwidth}{!}{
\begin{tabular}{l|cc}
\toprule
Method & IoU (\%) & mIoU (\%) \\
\midrule
VoxFormer-S~\cite{VoxFormer} & 38.68 & 10.67 \\
VoxFormer-T~\cite{VoxFormer} & 38.08 & 11.27 \\
Symphonize~\cite{Symphonize} & 38.37 & 12.20 \\
OccFormer~\cite{OccFormer} & 36.50 & 13.46 \\
CGFormer~\cite{yucontext} & {41.82} & {14.06} \\ \hline
\textbf{\Method (Ours)}  & \Best{43.92} & \Best{16.35} \\ 
\bottomrule
\end{tabular}
}
\label{tab:mono}
\end{minipage}
\end{table*}

\begin{SCfigure}[][!t]
\centering
\includegraphics[width=0.65\linewidth]{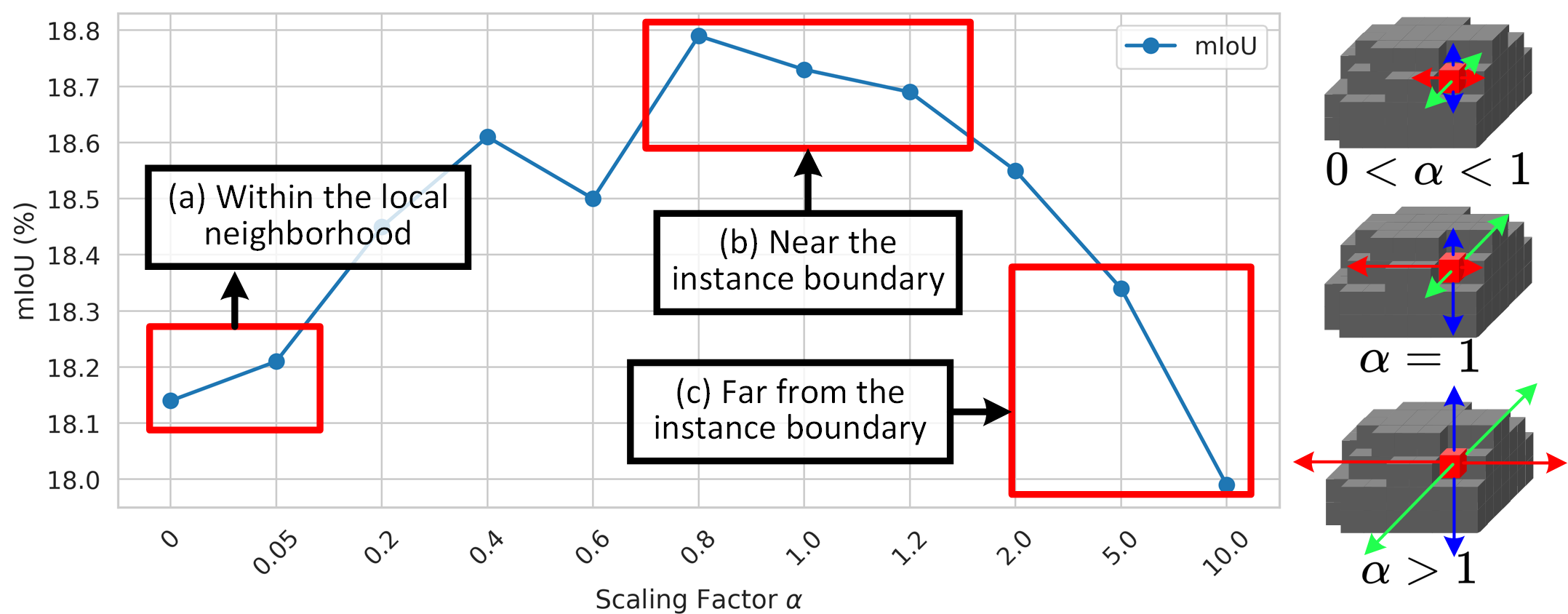}
\caption{Justifying the aggregation design using voxels at regressed positions $\Voxel_{\delta}$. By modulating regressed offsets $\Delta$ with a scaling factor $\alpha\in\mathbb{R}$, for each voxel $\Voxel_{i,j,k}$, we select the voxels $\Voxel_{\delta}$ based on modulated offsets $\Tilde{\Delta}=\alpha\,\Delta_{i,j,k}$, where increasing $\alpha$ will select voxels at larger distances and vice versa.\vspace{-8pt}}
\label{fig:agg}
\vspace{-8pt}
\end{SCfigure}

\noindent\textbf{Ablation Study.} Tab.~\ref{tab:ablation} presents the ablation study, highlighting the following insightful observations. \highlight{Line (b):} Compared with the baseline, introducing the REG branch in TDP gives a significant 2.71 IoU gains owing to the better instance-level perception. Interestingly, the limited 0.14 mIoU shows that the REG does not contribute to the semantic discrimability, aligning with our motivation. \highlight{Line (c):} After deploying the CLS with instance-level aggregation, a significant 1.43 mIoU gain can be observed, revealing the importance of instance guidance in voxel prediction. \highlight{Line (d-f):} By gradually decoupling the volume, both metrics improve progressively, eliminating the task misalignment. 

\noindent\textbf{Further Analysis.} In Tab.~\ref{tab:further}, we delve into each module with different design variants. For \textbf{TDP}, replacing the $\Delta$-guided instance-level aggregation (Eq.~\ref{eq:adaptive}) with learned deformable self-attention and soft-weights both decreases the performance, revealing the necessity of explicit instance guidance. \emph{The $\Delta$ guidance and attentive design may relieve the negative influence of regression outliers far from instances, consistent with Fig.~\ref{fig:agg}}. In \textbf{SVE}, we observe a consistent decline by replacing dense projection, $\text{DefConv}$, TPV, and decoupling with average pooling, $\Conv$, $\text{ResBlock}$, and task-shared designs, respectively. This can verify that our spatial task-decoupling in the TPV space is optimal.

\noindent\textbf{Model Efficiency.} In Tab.~\ref{tab:efficiency}, we compare with state-of-the-art works in parameters (N$_{\mathrm{param}}$) and inference time (T$_{\mathrm{inf}}$). \Method uses significantly fewer parameters (M) and less inference time (ms), setting new records on all IoU metrics, highlighting the effectiveness of our new formulation. 

\noindent\textbf{Comparison with Monocular Depth.} In Tab.~\ref{tab:mono}, we make a comparison with state-of-the-art methods by using monocular depth~\cite{Adabins}. Our \Method also achieves the best results on both metrics, surpassing the previous best entry with 2.10 IoU and 2.29 mIoU, showing significant robustness on depth.

\noindent\textbf{Exploring Instance-level Aggregation.} Fig.~\ref{fig:agg} carefully analyzes our aggregation designs (Eq.~\ref{eq:adaptive}). Instead of directly aggregating regressed voxels $\Voxel_{\delta}$ with $\Delta$, we modulate it by a scaling factor $\alpha \in \mathbb{R}$, and aggregate at the scaled offset $\Tilde{\Delta}=\alpha\Delta$, revealing three observations. \highlight{(1) Local aggregation does not work ($\alpha\!\to\!0$).} When $\alpha$ approaches zero, performance degrades noticeably. Naively aggregating neighborhoods yields no benefit, as it degenerates to convolutional kernels with local receptive fields. \highlight{(2) Near-boundary aggregation helps ($\alpha\!\approx\!1$).} As $\alpha$ grows toward 1.0, performance steadily improves, indicating that the boundary contributes more informative signals for dense perception. Notably, $\alpha=0.8$ slightly outperforms our default of $\alpha=1.0$, likely by suppressing outliers outside instances. \highlight{(3) Outside-instance aggregation hurts ($\alpha\!>\!1$).} When $\alpha$ exceeds 1.0 substantially, performance declines sharply, revealing the negative impact of aggregating voxels beyond the instance extent.

\begin{figure*}[t]
\centering\includegraphics[width=0.90\linewidth]{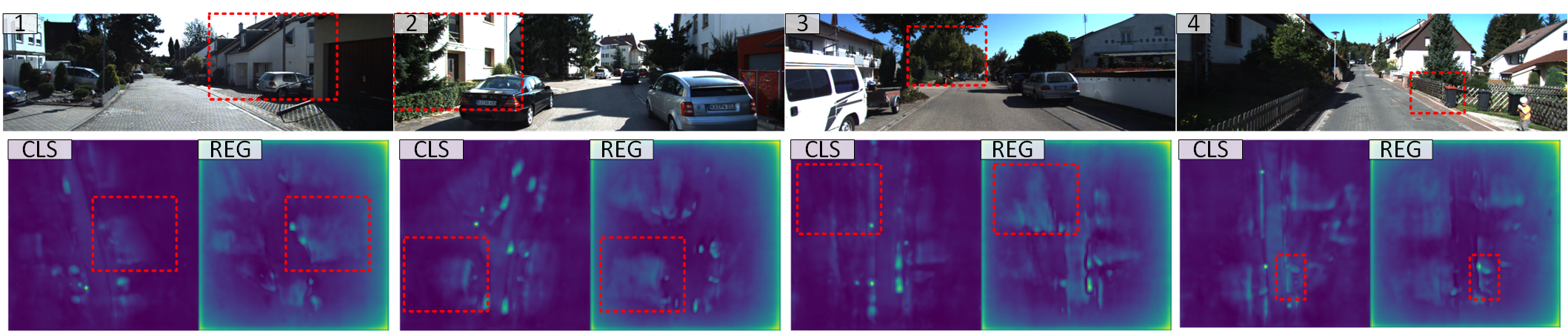}
\vspace{-5pt}
\caption{Visualization of the decoupled feature for classification (CLS) and regression (REG).}
\label{fig:feat}
\vspace{-10pt}
\end{figure*}

\begin{figure*}[t]
\centering\includegraphics[width=0.9\linewidth]{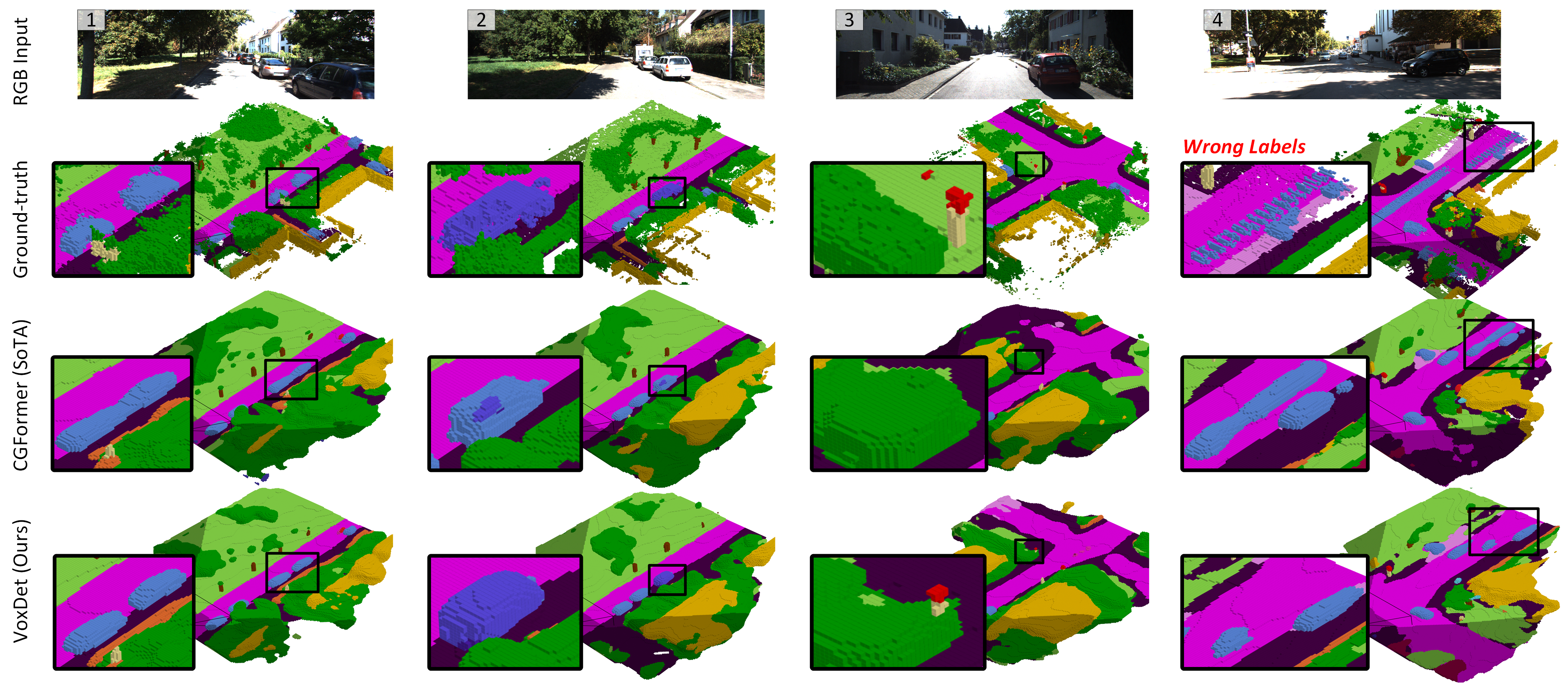}
\vspace{-4pt}
\caption{Qualitative comparison with state-of-the-art method~\cite{yucontext}. Zoom in for a better view.}
\label{fig:show}
\vspace{-13pt}
\end{figure*}

\subsection{Qualitative Study}

\noindent\textbf{Disentangled Volume.} In Fig.~\ref{fig:feat}, we visualize the decoupled feature volume (Eq.~\ref{eq:decouple}) at Bird's Eye View (BEV) view for the classification (CLS) and regression (REG), revealing the following three insightful observations. \textbf{(1)} Different from CLS, focusing on semantically informative local regions, REG naturally seeks to discover the more complete boundary of instances. This capacity is critical, as completing the scene is the key focus. \textbf{(2)} Although REG has not been trained with semantic supervision, it can discover informative cues of potential objects (4$^{\text{th}}$ sample) with orthogonal effects with CLS. \textbf{(3)} We can see the four borders of the BEV map are activated, as the 3D volume borders are unified boundaries of adjacent large-scale instances, like vegetation and buildings.

\noindent\textbf{Semantic Occupancy Prediction.} Fig.~\ref{fig:show} visually compares our \Method with state-of-the-art method~\cite{yucontext}. VoxDet shows noticeable gains in geometric completion, such as complete \emph{car} borders in the 1$^{\text{st}}$ sample, better infers instance-level semantics, such as the \emph{truck} in the 2$^{\text{nd}}$ sample, and can correctly detect the \emph{challenging yet important traffic sign} (3$^{\text{rd}}$ sample). This is mainly owing to the proposed instance-centric formulation, which enhances spatial perception and semantic understanding of object instances. Notably, in the 4$^{\text{th}}$ sample, instead of wrongly fitting wrong labels, VoxDet gives more reasonable predictions on dynamic objects, which is further discussed in the Appendix.

\section{Conclusion}
In this work, we first reveal an ever-overlooked free lunch in occupancy labels: the voxel-level class label has implicitly provided the instance-level insight thanks to its occlusion-free nature. Then, we propose \textbf{\Trick}, a simple yet effective algorithm that freely converts class labels to the instance-level offsets. Based on this, we propose \textbf{\Method}, a new detection-based formulation for instance-centric prediction. \Method first spatially decouples 3D volumes to generate task-specific features, learning spatial deformations in the tri-perceptive space. Next, it adopts a task-decoupled predictor to generate instance-aware predictions guided by the regressed 4D offset field. Extensive experiments reveal a lot of insightful phenomena for the following works and verify the state-of-the-art role of \Method.

\subsection{Acknowledgment}

We would like to express our gratitude to Reyhaneh Hosseininejad, Yasamin Borhani, Yuanfan Zheng, and Hengyu Liu for their insightful discussions, which contributed to the improvement of this work. Additionally, we gratefully acknowledge the financial support provided by Valeo.

\section*{Appendix}
\appendix

We provide comprehensive supplementary materials to clarify novelty, practical applications, border impacts, experimental justifications, and future directions, potentially inspiring the following works, which are organized as follows. \textbf{Bold} highlights the primary sections with more insights. 

\begin{enumerate}[label={}]
 \item Appendix~\ref{sec:quant_exp}: Additional quantitative results
 \begin{itemize}

 \item \textbf{Robust analysis with multiple runs}
 \begin{itemize}
 \item Qualified as a new, powerful, robust, lightweight, and efficient baseline
 \end{itemize}
 \item Results on SemanticKITTI validation set
 \item \textbf{How to define object instances}
 \begin{itemize}
 \item A generalized definition works best, which is different from the 2D intuition
 \end{itemize} 
 \item Sensitivity analysis on hyperparameters
 \item More comparison of model efficiency
 
 \end{itemize}

 \item Appendix~\ref{sec:trick_more}: \textbf{More insights and practical usages of VoxNT trick}
 \begin{itemize}
 \item Freely understand instance scales
 \item Ability to freely identify wrong labels
 \item Freely eliminate wrong labels in training
 \item Rethink the evaluation on dynamic objects
 
 \end{itemize} 
 \item Appendix~\ref{sec:exp_setup}: Detailed experimental setups
 \begin{itemize}
 \item Datasets and evaluation metrics
 \item Implementation details
 \item Algorithmic details of the \Trick trick
 \end{itemize} 

 \item Appendix~\ref{sec:discussion}: Discussion
 \begin{itemize}
 \item Difference with detection-assisted works
 \item \textbf{Broader impacts}
 \item \textbf{Limitations and future works}
 \item Ethical claims
 \end{itemize} 
 \item Appendix~\ref{sec:qual_exp}: Additional qualitative results
 \begin{itemize}
 \item Failure case analysis
 \item More qualitative comparison
 \end{itemize}
\end{enumerate}

\section{Additional Quantitative Results}\label{sec:quant_exp}

\subsection{Robustness Analysis of Multiple Runs} 

\begin{figure}[t]
\centering\includegraphics[width=0.9\linewidth]{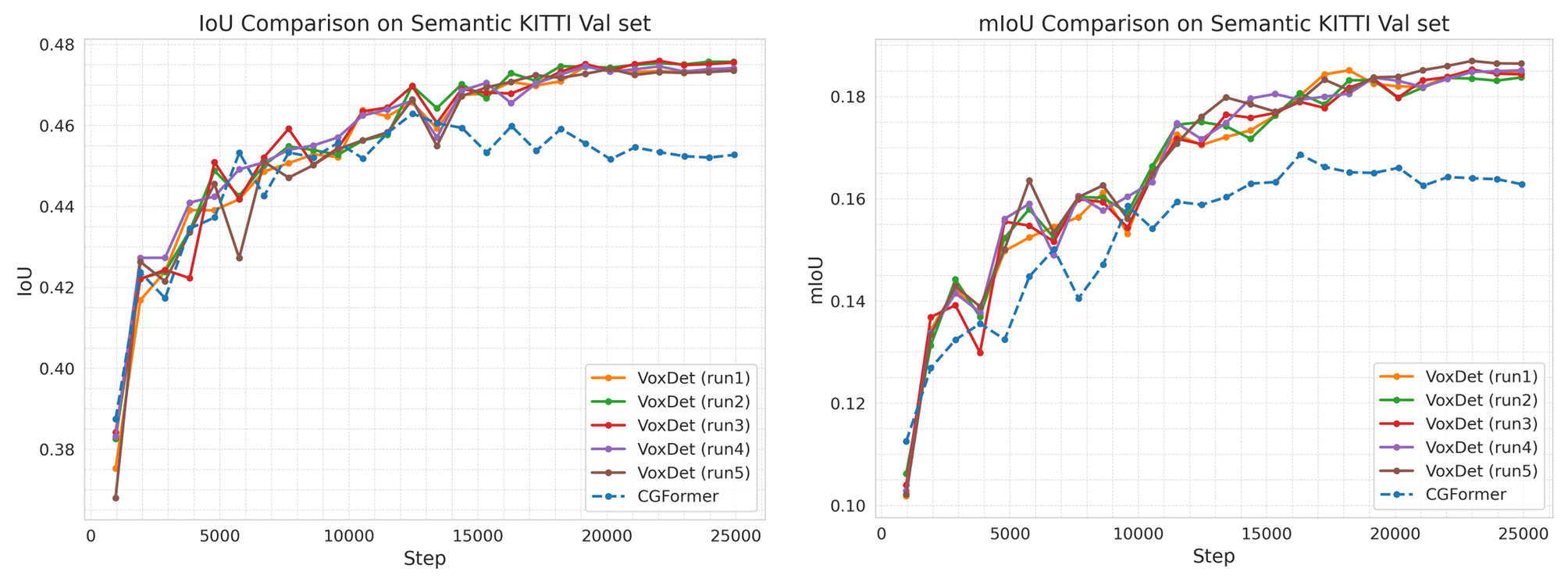}
\vspace{-0.5mm}
\caption{Robustness analysis of our VoxDet with multiple runs. We report the per-epoch validation results. The results from the official training log (not our reproduced results) given by the previous state-of-the-art method, CGFomrer~\cite{yucontext}, are also listed for comparison. We can observe visually significant improvements in both performance and robustness of our method.}
\label{fig:multi-run}
\vspace{-10pt}
\end{figure}

In Fig.~\ref{fig:multi-run}, we report the results of multiple runs (5 times independent experiments) to justify the robustness of our \Method. The curves summarize the IoU and mIoU results for each epoch on the SemanticKITTI validation set. To better illustrate the difference, we also demonstrate the results of the previous state-of-the-art method, CGFormer~\cite{yucontext}, using the officially released training log.

It can be observed that our VoxDet achieves impressive robustness with multiple runs, giving very similar performance on the IoU and mIoU in different runs. Note that there is some tradeoff between IoU and mIoU metrics, i.e., some runs achieve slightly higher IoU while sacrificing a little mIoU. Additionally, our method achieves visually significant gains over CGFormer on both the robustness and IoU/mIoU metrics, highlighting the strength \Method. \highlight{Hence, due to the superior effectiveness, robustness, and efficiency, we believe that \Method is qualified to serve as a powerful, lightweight, and efficient baseline model for the following works.}

\subsection{How to Define Object Instances?}

\begin{wraptable}{r}{7.0cm}
\setlength{\tabcolsep}{3pt}
\caption{\label{tab:obj} Analysis of object definitions by removing instance-level regression for specific classes.}
\resizebox{1.0\linewidth}{!}{
\begin{tabular}{l|C{2.2cm}C{2.2cm}}
\toprule
Setting & IoU & mIoU\\
\hline 
w/o. regress background & 46.78 & 18.19\\
w/o. regress empty & 47.08 & 18.05\\
\midrule
Full model & \Best{47.36} &\Best{18.73}\\
\bottomrule
\end{tabular}}
\end{wraptable}

In Tab.~\ref{tab:obj}, we explore different definitions of objects by removing the instance-level regression for specific classes, including (1) background, including road, sidewalk, etc (2) empty class. We empirically find that a generalized definition of objects, i.e., all the semantic classes, performs best. This is different from the intuition in the 2D image domain, which only considers objects with well-defined shapes.

\subsection{Results on SemanticKITTI Validation Set}

Tab.~\ref{tab:sem_kitti_val} presents the comparison on the SemanticKITTI validation set. Our \Method archives the best results of 47.36 IoU and 18.73 mIoU, surpassing the second-best counterpart~\cite{li2024hierarchical} with a noticeable 1.60 mIoU gains. Compared with the previous state-of-the-art method~\cite{yucontext} with 45.99 IoU and 16.87 mIoU, our \Method gives a 1.37 IoU and 1.86 mIoU gains, verifying the superior effectiveness. Besides, \Method achieves the best and second-best results in 17 of the 19 classes, which shows its effectiveness. Specifically, our method performs well on object instances, achieving the best on \emph{building, car, truck, motorcycle, traffic sign, etc}, revealing the superior instance-centric learning.

\begin{table*}
\newcommand{\clsname}[2]{
  \rotatebox{90}{
    \hspace{-6pt}
    \textcolor{#2}{$\blacksquare$}
    \hspace{-6pt}
    \renewcommand\arraystretch{0.6}
    \begin{tabular}{l}
      #1                                      \\
      \hspace{-4pt} ~\tiny(\semkitfreq{#2}\%) \\
    \end{tabular}
  }
}
\centering
\caption{Quantitative results on SemanticKITTI~\cite{SemanticKITTI} validation set. T indicates using extra temporal information. The best and the second best results are in  {bold} and  {underlined}, respectively.}\vspace{+3pt}
\resizebox{\linewidth}{!}{
\begin{tabular}{l|c|cc|ccccccccccccccccccc}
\toprule
Method                                 &
 {T}                                 &
IoU                                    &
mIoU                                   &
\clsname{road}{road}                  &
\clsname{sidewalk}{sidewalk}          &
\clsname{parking}{parking}            &
\clsname{other-grnd.}{otherground}    &
\clsname{building}{building}          &
\clsname{car}{car}                    &
\clsname{truck}{truck}                &
\clsname{bicycle}{bicycle}            &
\clsname{motorcycle}{motorcycle}      &
\clsname{other-veh.}{othervehicle}    &
\clsname{vegetation}{vegetation}      &
\clsname{trunk}{trunk}                &
\clsname{terrain}{terrain}            &
\clsname{person}{person}              &
\clsname{bicyclist}{bicyclist}        &
\clsname{motorcyclist}{motorcyclist}  &
\clsname{fence}{fence}                &
\clsname{pole}{pole}                  &
\clsname{traf.-sign}{trafficsign}     \\
\midrule
MonoScene~\cite{MonoScene}    &  & 36.86 & 11.08 & 56.52 & 26.72 & 14.27 & 0.46  & 14.09 & 23.26 & 6.98  & 0.61  & 0.45  & 1.48  & 17.89 & 2.81  & 29.64 & 1.86  & 1.20  & 0.00  & 5.84  & 4.14  & 2.25 \\
TPVFormer~\cite{TPVFormer}           &  & 35.61 & 11.36 & 56.50 & 25.87 & 20.60 & 0.85  & 13.88 & 23.81 & 8.08  & 0.36  & 0.05  & 4.35  & 16.92 & 2.26  & 30.38 & 0.51  & 0.89  & 0.00  & 5.94  & 3.14  & 1.52 \\
OccFormer~\cite{OccFormer}           &  & 36.50 & 13.46 &  {58.85} & 26.88 & 19.61 & 0.31  & 14.40 & 25.09 &  {25.53} & 0.81  & 1.19  &  {8.52}  & 19.63 & 3.93  & 32.62 & 2.78  & 2.82  & 0.00  & 5.61  & 4.26  & 2.86 \\
IAMSSC~\cite{IAMSSC}                 &  & 44.29 & 12.45 & 54.55 & 25.85 & 16.02 & 0.70  & 17.38 & 26.26 & 8.74  & 0.60  & 0.15  & 5.06  & 24.63 & 4.95  & 30.13 & 1.32  & 3.46  & 0.01  & 6.86  & 6.35  & 3.56 \\
VoxFormer~\cite{VoxFormer}         &  & 44.02 & 12.35 & 54.76 & 26.35 & 15.50 & 0.70  & 17.65 & 25.79 & 5.63  & 0.59  & 0.51  & 3.77  & 24.39 & 5.08  & 29.96 & 1.78  & 3.32  & 0.00  & 7.64  & 7.11  & 4.18 \\
VoxFormer~\cite{VoxFormer}         & \checkmark & 44.15 & 13.35 & 53.57 & 26.52 & 19.69 & 0.42 & 19.54 & 26.54 & 7.26 & 1.28 & 0.56 & 7.81 & 26.10 & 6.10 & 33.06 & 1.93 & 1.97 & 0.00 & 7.31 & 9.15 & 4.94 \\
Symphonize~\cite{Symphonize}        &   & 41.92 &  {14.89} & 56.37 & 27.58 & 15.28 & \SecondBest{0.95}  & 21.64 & 28.68 &  \SecondBest{20.44} &  {2.54}  &  \SecondBest{2.82}  &  \SecondBest{13.89} & 25.72 & 6.60  & 30.87 &  {3.52}  & 2.24  & 0.00  & 8.40  & 9.57  & 5.76 \\
HASSC~\cite{HASSC}               &    & 44.82 & 13.48 & 57.05 & 28.25 & 15.90 &  \Best{1.04} & 19.05 & 27.23 & 9.91 & 0.92 & 0.86 & 5.61 & 25.48 & 6.15 & 32.94 &  \SecondBest{2.80} & \Best{4.71} & 0.00 & 6.58 & 7.68 & 4.05 \\
H2GFormer~\cite{H2GFormer}     &      & 44.57 & 13.73 & 56.08 & 29.12 & 17.83 & 0.45 & 19.74 & 28.21 & 10.00 & 0.50 & 0.47 & 7.39 & 26.25 & 6.80 & 34.42 & 1.54 & 2.88 & 0.00 & 7.24 & 7.88 & 4.68 \\
H2GFormer~\cite{H2GFormer}     &   \checkmark   & 44.69 & 14.29 & 57.00 &  {29.37} &  \SecondBest{21.74} & 0.34 & 20.51 & 28.21 & 6.80 & 0.95 & 0.91 &  {9.32} &  \SecondBest{27.44} & 7.80 &  {36.26} & 1.15 & 0.10 & 0.00 & 7.98 &  {9.88} & 5.81 \\
SGN~\cite{SGN}&& 43.60&14.55 &59.32&30.51&18.46&0.42&21.43&31.88&13.18&0.58&0.17&5.68&25.98&7.43&34.42&1.28&1.49&0.00&9.66&9.83&4.71\\
SGN~\cite{SGN}&\checkmark& \SecondBest{46.21}&15.32&59.10&29.41&19.05&0.33&\SecondBest{25.17}&33.31&6.03&0.61&0.46&9.84&\Best{28.93}&\SecondBest{9.58}&38.12&0.47&0.10&0.00&9.96&13.25&7.32\\
CGFormer~\cite{yucontext} &  &  {45.99} &  {16.87} &  \SecondBest{65.51} &  {32.31} & 20.82 & 0.16  &  {23.52} &  \SecondBest{34.32} & 19.44 &  \Best{4.61} &  {2.71}  & 7.67  &  {26.93} &  {8.83}  &  \SecondBest{39.54} & 2.38  & \SecondBest{4.08}  & 0.00  &  {9.20} &  {10.67} &  \SecondBest{7.84} \\
HTCL~\cite{li2024hierarchical}& \checkmark & 45.51&\SecondBest{17.13}&63.70&\SecondBest{32.48}&\Best{23.27}&0.14&24.13&34.30&20.72&\SecondBest{3.99}&2.80&11.99&26.96&8.79&37.73&2.56&2.30&0.00&\SecondBest{11.22}&\SecondBest{11.49}&6.95\\
\midrule
\textbf{\Method (Ours)}&& \Best{47.36} &\Best{18.73} & \Best{65.55} & \Best{34.22} & {20.88} & 0.04 & \Best{25.79} & \Best{34.50} & \Best{31.05} & 3.95 & \Best{5.14} & \Best{14.65} & \Best{28.93} & \Best{10.20} & \Best{41.27} & \Best{4.48} & {3.14} & \SecondBest{0.00} & \Best{11.73} & \Best{12.19} & \Best{8.28} \\

\bottomrule
\end{tabular}
}
\setlength{\abovecaptionskip}{0cm}
\setlength{\belowcaptionskip}{0cm}
\label{tab:sem_kitti_val}
\end{table*}

\subsection{Sensitivity Analysis }
In this section, we further give a detailed analysis of the hyperparameters used in \Method. The experiments are conducted with the same random seed to ensure fair comparison. The results are reported on the SemanticKITTI validation set as the unified practice. 

\noindent\textbf{Number of Instance‐level Aggregation Layers.} Tab.~\ref{tab:sen_n} reports the performance changes as we vary the number of aggregation layers $N$. Performance improves steadily with increasing $N$, confirming the benefit of more effective aggregation. Although extending to five layers (one more than our default $N=4$) yields a slight additional gain, the marginal improvement will lead to extra computational overhead. We therefore adopt $N=4$ as our default setting.

\noindent\textbf{Weight of Auxiliary Loss.} Tab.~\ref{tab:sen_l} presents the sensitivity of our model to the auxiliary segmentation loss weight $\lambda$ in $\mathcal{L}^{\text{aux}}_{\text{occ}}$, which governs the strength of voxel-level supervision. Overall, mIoU remains relatively stable across a wide range of $\lambda$ values: as we increase $\lambda$, performance steadily improves, peaking at $\lambda = 0.2$, and then gradually declines for larger weights. This trend indicates that a moderate auxiliary loss provides beneficial guidance for voxel-wise feature learning, while an overly large weight interferes with the subsequent instance-centric optimization. Consequently, we adopt $\lambda = 0.2$ as our default setting, as it yields the best mIoU.

\noindent\textbf{Loss weight of Regression.} In Tab.~\ref{tab:sen_r}, we add a loss weight on the regression $\lambda_{\mathrm{reg}}$ and explore the effect of varying the loss weight. A moderate increase in $\lambda_{\mathrm{reg}}$ yields further IoU gains, likely due to improved fitting of instance contours. However, excessively large weights cause a slight mIoU drop, probably from conflicts among the combined loss terms. We therefore set $\lambda_{\mathrm{reg}} = 1.0$ to balance the effect and eliminate the term in the main paper for convenience.
\begin{table*}[t]
\footnotesize
\centering
\begin{minipage}[t]{0.32\linewidth}
\centering
\caption{Analysis on instance-level aggregation layers $N$.}
\resizebox{0.6\textwidth}{!}{
\begin{tabular}{c|cc}
\toprule
 $N$ & IoU & mIoU \\\midrule
1 & 47.06 &  18.20\\
2 &  \Best{47.38}&   18.32\\
3 &  {47.34}&   18.54\\
4 &{47.36}  & {18.73} \\
5 &\Best{47.39}  & \Best{18.79} \\
\bottomrule
\end{tabular}}
\label{tab:sen_n}
\end{minipage}
\hfill
\begin{minipage}[t]{0.32\linewidth}
\centering
\caption{Analysis on the loss weight $\lambda$ deployed on $\mathcal{L}^{\text{aux}}_{\text{occ}}$.}
\resizebox{0.6\textwidth}{!}{
\begin{tabular}{c|cc}
\toprule
 $\lambda$ & IoU & mIoU \\\midrule
0.1 &{47.15}  & {18.53} \\
0.2 &{47.36}  &\Best{18.73} \\ 
0.4 &\Best{47.43}  & {18.69} \\
0.8 &47.28  & {18.59} \\
1.0 &{47.39}  & {18.42} \\
\bottomrule
\end{tabular}}
\label{tab:sen_l}
\end{minipage}
\hfill
\begin{minipage}[t]{0.32\linewidth}
\centering
\caption{Analysis on the loss weight $\lambda_{\text{reg}}$ deployed on $\mathcal{L}_{\text{Reg}}$ .}
\resizebox{0.6\textwidth}{!}{
\begin{tabular}{c|cc}
\toprule
 $\lambda_{\text{reg}}$ & IoU & mIoU \\\midrule
0.5 & 47.19  & 18.46  \\
1.0 &{47.36}  & \Best{18.73} \\
1.5 & 47.48 &  18.68\\
2.0 & 47.55  & 18.59 \\
2.5 &  \Best{47.68}&  18.55\\
\bottomrule
\end{tabular}}
\label{tab:sen_r}
\end{minipage}
\vspace{-10pt}
\end{table*}

\subsection{Model Efficiency}

 Tab~\ref{tab:time} presents a comprehensive comparison between our \Method and current state‐of‐the‐art methods on the SemanticKITTI test set, evaluating model size, inference speed, and prediction performance. \Method requires the fewest learnable parameters, achieves the fastest inference time, and attains the highest per‐class IoU as well as the best overall mIoU. Compared to its strongest competitor~\cite{bae2024three}, our approach slashes the parameter count, accelerates inference to real‐time performance, and delivers a notable gain in mIoU. These results underscore the strength of our effective formulation, giving a lightweight yet powerful framework that simultaneously advances speed, efficiency, and accuracy.

\begin{table}[t]
\centering
\caption{Comparison of model efficiency and accuracy with SoTA on SemanticKITTI test set.}\vspace{+5pt}
\resizebox{\linewidth}{!}
{
 \begin{tabular}{l|lllllll|l}
 \toprule
 Venue & CVPR' 23& CVPR' 23& CVPR' 23& ICCV'23 &IJCAI' 24 & NeurIPS' 24 & CVPR' 25 \\
 Method & TPVFormer~\cite{TPVFormer} & VoxFormer~\cite{VoxFormer} & Symphonize~\cite{Symphonize}&OccFormer~\cite{OccFormer} & StereoScene~\cite{StereoScene} & CGFormer~\cite{yucontext} & ScanSSC~\cite{bae2024three} & \textbf{\Method}\\ 
 \midrule
 Param. (M) $\downarrow$&107 & 59 & 59&214 &117 & 122 & 145 & \Best{53}\\
 Inf. Time (ms) $\downarrow$ & 207 & 204 & 216 & 199 & 258 & 205 & 261& \Best{159} \\\midrule
 IoU $\uparrow$ & 34.25 & 42.95 & 42.19 &34.53 &43.34 & {44.41}& 44.54& \Best{47.27} \\
 mIoU $\uparrow$ & 11.26 & 12.20 & 15.04 &12.20 &15.36 & {16.63} & 17.40 & \Best{18.47}\\
 \bottomrule
 \end{tabular}
}\label{tab:time}
\end{table}

\section{Voxel-to-Instance (VoxNT) Trick}\label{sec:trick_more}

\begin{figure*}[t]
\centering\includegraphics[width=1.0\linewidth]{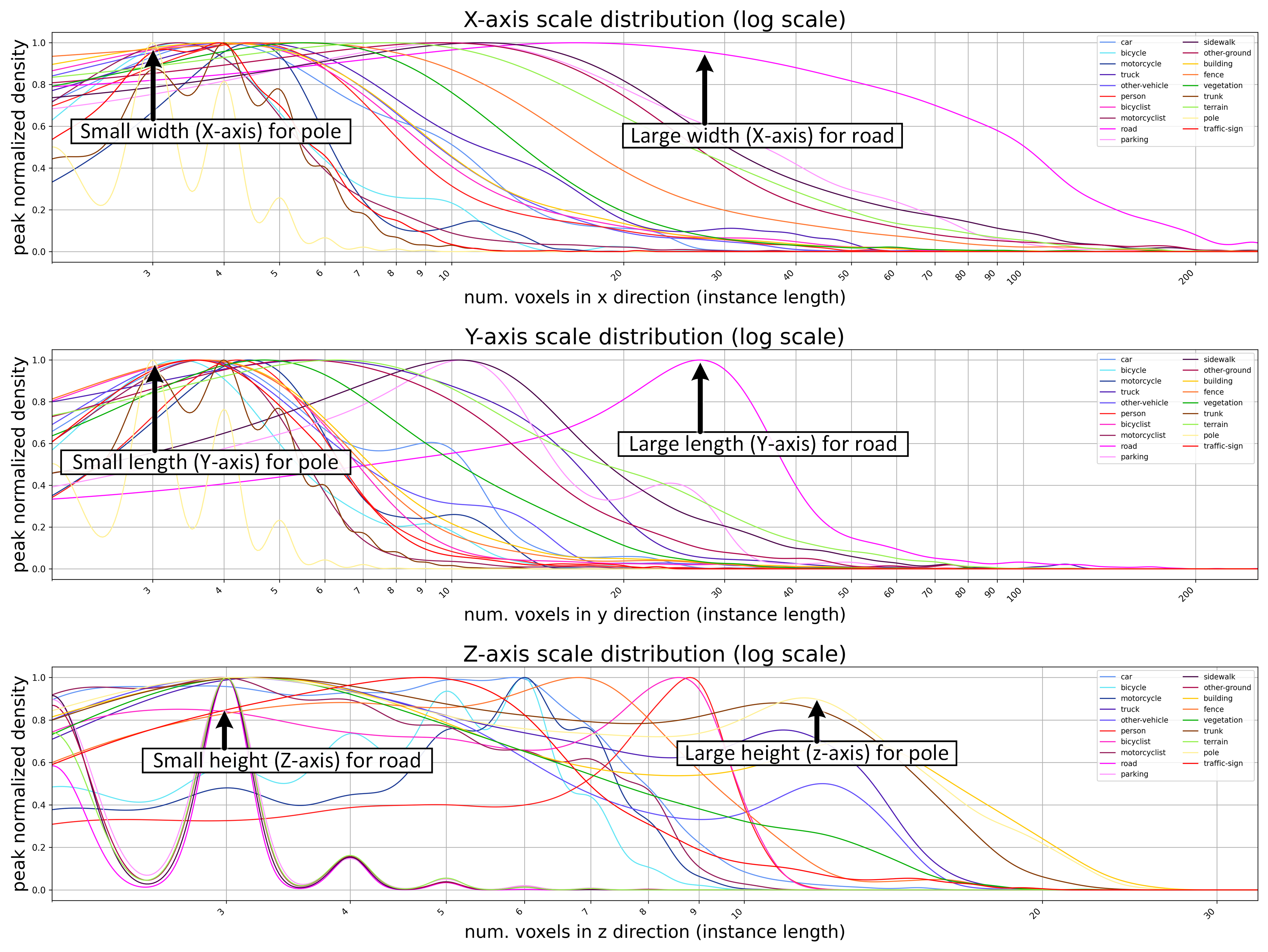}
\vspace{-4pt}
\caption{Instance-level scale distribution in $X,Y,Z$ axis given by our voxel-to-instance trick. We randomly sample voxels in all classes and calculate the instance scale ($l$) with coupled offset terms (positive and negative directions), e.g., $l^x = \delta^{x^+}+\delta^{x^-}$ in $X$-axis. The horizontal axis represents the scale (the number of voxels) on a log scale. The scale of the whole scene in $X, Y, Z$ is $256,256,32$ respectively. The vertical represents the number of samples, which is peak normalized to $[0,1]$ by dividing the maximum number of each class for a better view. Zoom in for a better view.}
\label{fig:offset_distribution}
\vspace{-2mm}
\end{figure*}

\subsection{VoxNT Trick Can Freely Understand Instance Scales} 

In Fig.~\ref{fig:offset_distribution}, we conduct a statistical analysis on the instance scale along different axes based on the proposed Voxel-to-Instance (VoxNT) Trick. To better visualize the scale patterns, we visualize the distribution for different classes (in different colors), highlighting their specific scale patterns. The resolution of the whole scene is $256\times256\times32$. To better explore the instance-level geometries, we sum up the positive and negative directions of the 4D offset field $\Delta$ to represent the scale information in different axes. This process is denoted as follows,
\begin{equation}
 \begin{split}
 l^x &= \delta^{x^+}+\delta^{x^-}; \\
 l^y &= \delta^{y^+}+\delta^{y^-}; \\
 l^z &= \delta^{z^+}+\delta^{z^-}. \\
 \end{split}
\end{equation}
Here $\{\delta^{x^+},\delta^{x^-},\delta^{y^+},\delta^{x^-},\delta^{z^+},\delta^{z^-}\}$ indicates the offset in the six directions. Surprisingly, we find that $\Delta$ is able to freely give sufficient instance-level cues beyond the spatially agnostic semantic labels: the essential scale information. Besides, the geometric scale distribution of different classes also demonstrates different unique patterns, greatly aligning with the real-world environment. The following are intuitive observations.

\textbf{(1) For tall instances}, such as \emph{pole} in the light yellow color, the scales tend to give small values in $X$ and $Y$ axes (usually lower than six voxel width/length), and provide a large scale on the $Z$ axis (e.g., larger than 10-voxel height.). This is aligned with the comment scene as the width and length of these things are small, while the height is significant.

\textbf{(2) For the flat items}, such as \emph{road} in the pink color, we can see a significant value in $X$ and $Y$ axes (usually larger than 30 voxel width/length), while giving a small scale on the $Z$ axis (smaller than 5 voxels), which also aligns with our intuition.

\textbf{(3) For some large objects}, such as \emph{vegetation} in the green color, we can see that the scale in all three axes can be relatively significant, aligning with the real-world environment.

Hence, these observations further highlight the value of the proposed VoxNT trick, which can convey extensive valuable information in geometry at the instance level. Note that this information is not available in the original voxel-level class labels.

\begin{figure*}[t]
\centering\includegraphics[width=0.9\linewidth]{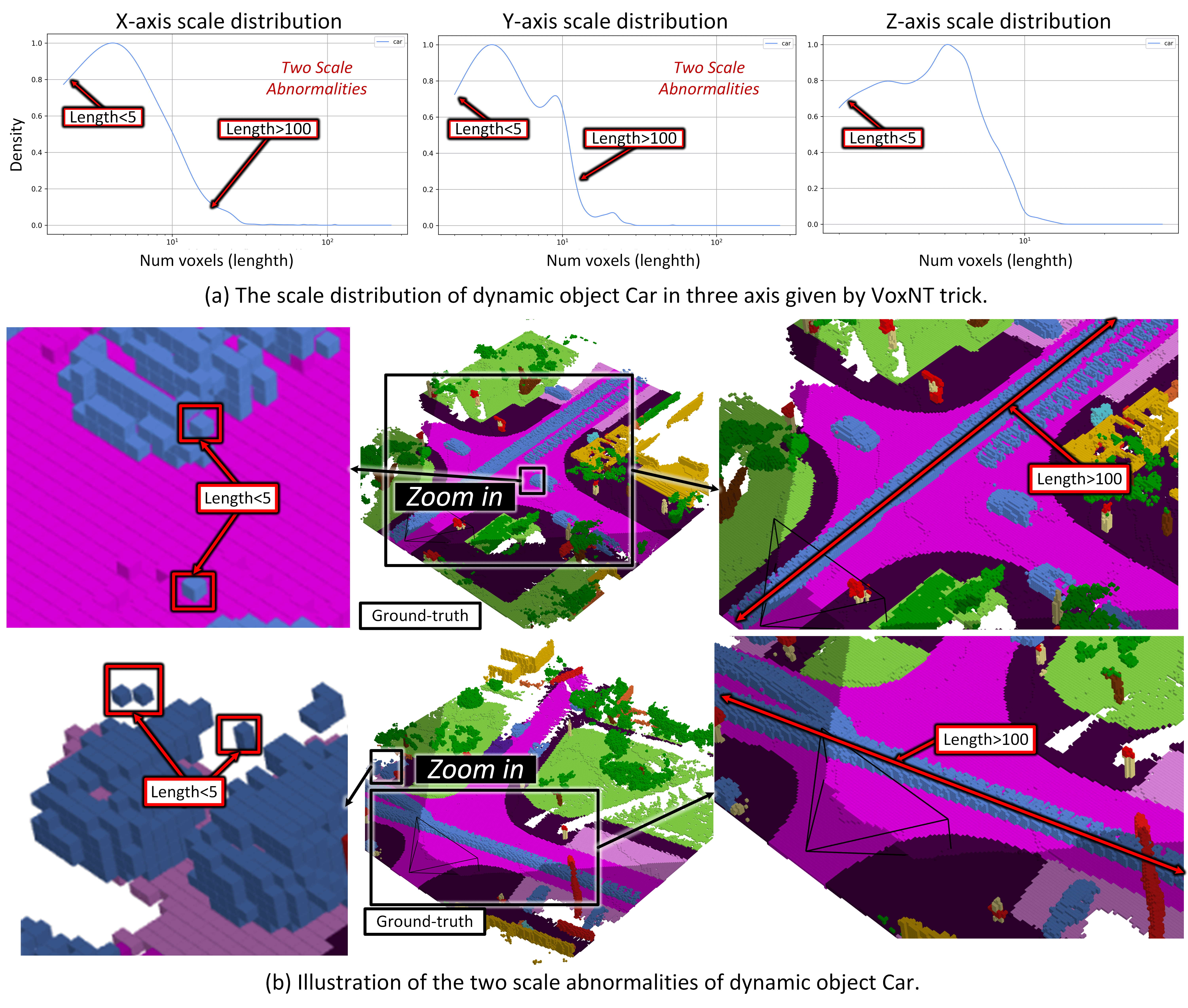}
\vspace{-0.5mm}
\caption{More observations from VoxNT. (a) Scale distribution (same as App.~\ref{sec:trick_more}) of the car along the $X,Y,Z$ axis. \textbf{We can see two types of abnormality: minimal and large lengths.} (b) Visualizting the abnormality with ground-truth. The left shows the isolated voxels leading to the minimal-scale abnormality. The right shows the large-length abnormality caused by the traces of moving objects.}
\label{fig:car}
\vspace{-3mm}
\end{figure*}

\subsection{VoxNT Trick Can Freely Identify Wrong Labels}

\textbf{Observe the Scale Abnormality of Dynamic Objects with VoxNT Trick.} We first study the \emph{car} class\footnote{Cars serve as the key target and dominate the evaluation of detection benchmarks~\cite{chen2018domain,wang2021fcos3d}}, the most prevalent and dynamic object category in autonomous driving. Fig.~\ref{fig:car} (a) demonstrates the distribution of the instance scale along different axes, generated by our VoxNT trick. Given the whole scale of $256\times256\times32$ in the scene, we find that the scale of the car is problematic: In the $X$ and $Y$ axis, there are lots of samples with tiny scale with \textbf{length less than 5} and huge scale with \textbf{larger than 100}. In the $Z$ axis, many samples have \textbf{length less than 5}. Note that the typical scale of a car is less than $30\times20\times10$, indicating that the ground-truth has significant abnormality.

\noindent\textbf{Delve into the Abnormality with Label Visualization.} To study the essential reason, we visualize the ground-truth voxels in Fig.~\ref{fig:car} (b). We can ground the observed issues mentioned above according to the visualized ground truth, which is voxelized from point clouds. In the left sub-figure, we find that the reason for the samples with abnormally small length is \textbf{isolated voxels}. In the right sub-figure, we can see that the failed filter object dynamics lead to abnormally large lengths.

\subsection{VoxNT Trick Can Freely Eliminate the Influence of Wrong Labels in Training}

We can see that the scale information in our 4D offset field has a valuable property for solving this issue, which, for the first time, makes the abnormality identification realistic. In this section, we discuss some tricky uses of the proposed VoxNT trick to inspire the follow-up works. Note that these operations improve a more reasonable prediction but cannot improve the mIoU evaluation, because the ground-truth is noisy (see App.~\ref {sec:evaluate}). In brief, the key idea is to use scale information to identify the voxel with an offset that is too large or too small. Specifically, to identify the impact of the isolated voxels, we can use a binary voxel-level mask $\mathbf{M}^{\text{min}}_{i,j,k}\in \{0,1\}^{X\times Y\times Z}$ to \textbf{identify the voxels that have abnormally small offsets in all axes}:
\begin{equation}
\mathbf{M}^{\text{min}}_{i,j,k}=
 \begin{cases}
 1.0, & (l^x_{i,j,k} < K^{x}_{\text{min}}) \cap (l^y_{i,j,k} < K^{y}_{\text{min}}) \cap (l^y_{i,j,k} < K^{z}_{\text{min}}); \\
 0.0, & \text{otherwise},
 \end{cases}
\end{equation}
where $ K_{\text{min}} \in \mathbb{N}^{3}$ is a scale threshold for the three axes for measuring the removed minimal scale and can be empirically set to $3$. Similarly, to filter out the huge-scale samples, e.g., the car in Fig.~\ref{fig:car}, a similar mask can be deployed $\mathbf{M}^{\text{max}}_{i,j,k}\in \{0,1\}^{X\times Y\times Z}$ to \textbf{identify the voxels with abnormally large offsets appearing in one of the axes}. This can be written as follows,
\begin{equation}
\mathbf{M}^{\text{max}}_{i,j,k}=
 \begin{cases}
 1.0, & (l^x_{i,j,k} \geq K^x_{\text{max}} )\cup (l^y_{i,j,k} \geq K^y_{\text{max}})\cup (l^y_{i,j,k} \geq K^z_{\text{max}}) \\
 0.0, & \text{otherwise},
 \end{cases}
\end{equation}
where $K_{\text{max}}$ is the threshold filtering out the large lengths. Another class-based mask can also be adopted $\mathbf{M}^{\text{max}}_{i,j,k}$ with class labels ${Y}_{i,j,k}$ and desired class $K_c$ for the wrong label filtering, such as car. 

By using these two types of masks, we can clearly identify and localize those wrong voxels with the specific positions $(i,j,k)$ in the volumes. To remove their influence on training, it is possible to directly refine the ground-truth labels by ignoring these voxels with a straightforward yet effective label transformation. In practice, it can be implemented as the following equations,
\begin{equation}
 \label{eq:gt_refine}Y^{\text{refined}}_{i,j,k}= 
\begin{cases}
255, & (\mathbf{M}^{\text{max}}_{i,j,k} =1) \cup( \mathbf{M}^{\text{min}}_{i,j,k} =1 ) \cap( Y_{i,j,k}=K_{car});\\
K_{car}, & (\mathbf{M}^{\text{max}}_{i,j,k} =0 )\cap (\mathbf{M}^{\text{min}}_{i,j,k} =0) \cap( Y_{i,j,k}=K_{car}); \\
Y_{i,j,k}, & \text{otherwise}.
\end{cases}
\end{equation}
Here, the key idea is to set the label as ignored (255) if the scale of the car voxel is too large or too small. This can effectively remove lots of wrong predictions in the dynamic car.

\begin{figure}[t]
\centering\includegraphics[width=0.9\linewidth]{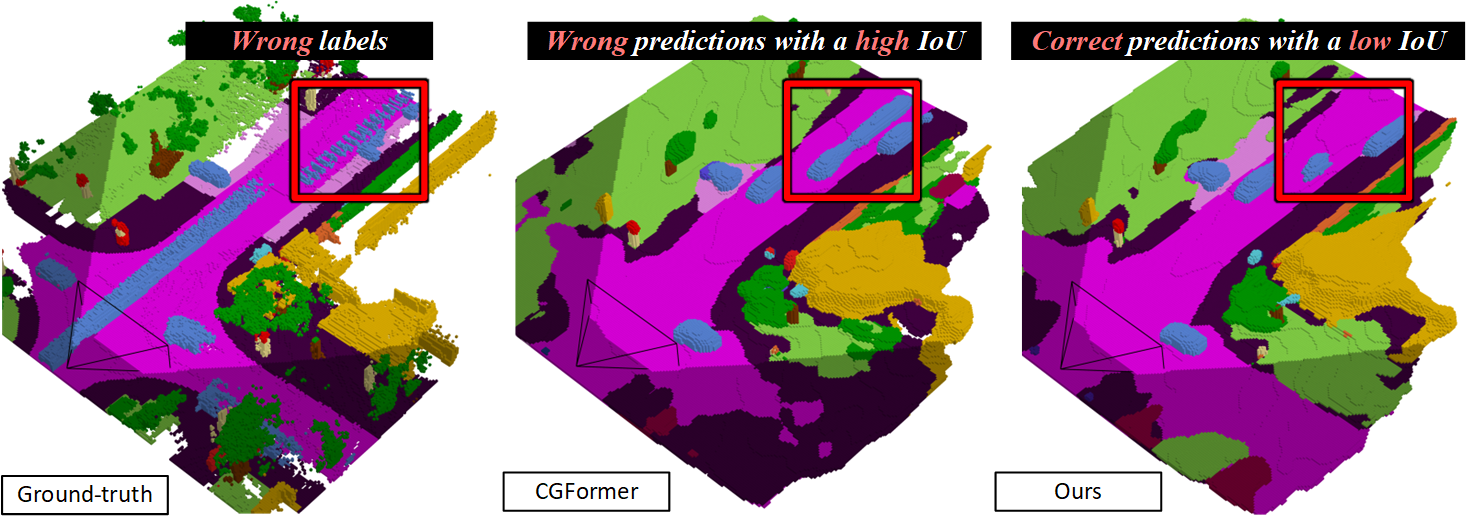}
\vspace{-0.5mm}
\caption{Illustration of the problems in the existing evaluation metrics. \textbf{Left:} The ground-truth wrongly labels the car with a sequential afterimage. \textbf{Middle:} Previous works wrongly fit these errors by predicting a long car. \textbf{Right:} Differently, our \Method can give more reasonable predictions without fitting the error, but the car IoU in this sample is worse due to the wrong labels.}
\label{fig:eval}
\vspace{-3mm}
\end{figure}
\subsection{Rethink the Evaluation on Dynamic Objects}\label{sec:evaluate}
Based on the observation about wrong labels, we further delve into the evaluation on the dynamic object \emph{car}. In Fig.~\ref{fig:eval}, we visualize the ground-truth of the sample from SemanticKITTI~\cite{SemanticKITTI} validation (Left), the corresponding prediction from the state-of-the-art method~\cite{yucontext} (Middle), and the prediction from our method (Right). We can see that previous works are prone to overfitting these wrong labels while generating higher results on the IoU metric for the car. Differently, our method gives a more reasonable prediction. However, this advantage cannot be demonstrated by using the conventional IoU metric due to the wrong labels. We hope this observation can inspire the following works to notice and address this issue, thereby pushing forward the community.

\section{Experimental Settings}\label{sec:exp_setup}

\subsection{Datasets and Metrics}

\noindent\textbf{Benchmark Setting.} Following the unified setting in this community, we evaluate our approach on two benchmarks: SemanticKITTI~\cite{behley2019semantickitti} and SSCBench‐KITTI‐360~\cite{li2024sscbenchlargescale3dsemantic}, which are derived from the original KITTI Odometry~\cite{geiger2012we} and KITTI‐360~\cite{liao2022kitti} datasets, respectively. In all experiments, we follow the unified setup~\cite{VoxFormer,yucontext,Symphonize}, restricting to the frustum of size $51.2\,$m in the forward direction, $\pm25.6\,$m laterally, and $6.4\,$m vertically above the sensor. This volume is voxelized into a grid of size $256\times256\times32$, with each voxel measuring $0.2\,$m on a side. Details are as follows. 

{(1) SemanticKITTI} consists of 22 sequences (00–21) of LiDAR scans and synchronized stereo images. We adopt the standard split of 10 sequences for training (00–07, 09–10), one sequence for validation (08), and 11 sequences for online evaluation on the hidden test server (11–21). Input images are provided at $1226\times370$ resolution, and ground‐truth occupancy grids are annotated with 20 labels (19 semantic classes + 1 empty class). 

{(2) SSCBench‐KITTI‐360} is a recently released extension that re‐labels a subset of the KITTI-360 sequences. It comprises 9 sequences in total, of which 7 are used for training, 1 for validation, and 1 held out for final testing. RGB images are captured at $1408\times376$ resolution, and each voxel is annotated with one of 19 labels (18 semantic categories + 1 free‐space class).

\noindent\textbf{Evaluation Metrics.} We assess performance along two complementary axes, i.e., geometry completion and semantic completion, using the Intersection over Union (IoU) and mean Intersection over Union (mIoU) metrics, for occupied voxel grids and voxel-wise semantic predictions. All reported results follow the evaluation protocols from previous works, including the online test‐server evaluations for SemanticKITTI~\cite{behley2019semantickitti}, ensuring a fair comparison with prior methods~\cite{VoxFormer,TPVFormer,yucontext}.

\subsection{Implementation Details}

\noindent\textbf{Backbone Network.} Following the main stream of occupancy prediction works~\cite{VoxFormer,MonoOcc}, we use ResNet-50~\cite{ResNet} as the 2D image feature extractor. We follow the unified SSC settings in the recent 2-year publications for the depth estimation and use MobileStereoNet~\cite{MobileStereoNet} and Adabins~\cite{Adabins} as the depth estimators. Note that the compared methods use the same depth images for the fair comparison. 

\noindent\textbf{View Transformation.} In the main paper, the 2D-to-3D view transformation follows~\cite{yucontext,bae2024three}. Specifically, given the input image $\mathbf{I}$, we first extract 2D image feature $\mathbf{F}^{\mathrm{2D}}$ and depth map $\mathbf{Z}$. Then, we adopt a depth refinement module that takes the 2D feature $\mathbf{F}^{\mathrm{2D}}$ and depth $\mathbf{Z}$ as input. 
It first estimates the depth distribution via LSS~\cite{lss}, and then generates voxel queries $\Voxel_{\mathbf{Q}} \in \mathbb{R}^{X\times Y \times Z \times C}$. Here, $(X,Y,Z)$ is spatial resolution $128 \times 128 \times 16$, and $C=128$ is the channel. 

Based on this, the pixel $(u,v)$ can be transformed to the 3D point $(x,y,z)$ using the camera intrinsic matrix ($\in \mathbb{R}^{4\times 4}$) and extrinsic matrix ($\in \mathbb{R}^{4\times 4}$), termed the query proposals $\mathbf{Q}$ in the projected voxel space. Finally, the 3D deformable cross-attention is deployed to query the information from 2D image to the 3D voxel space. The number of deformable cross-attention layers is $3$, and the number of sampling points around each reference point is set to $8$. This can generate the 3D feature volume with dimensions of $128 \times 128 \times 16$ and $128$ channels, which is the $\mathbf{V}$ in the main paper. For fair comparison, these operations in view transformation are the same as previous works~\cite{yucontext,bae2024three,wang2025l2cocc}. Kindly refer to~\cite{yucontext} for more details.

\noindent\textbf{\Method.} The number of instance-driven aggregation layers $N$ is set to 4. The loss weight terms of $\lambda$ and $\beta$ are empirically set to 1.0 and 0.2, respectively. For the task-shared voxel encoder in our spatially-decoupled voxel encoder, we directly use the encoder part of the conventional 3D UNet in other works~\cite{yucontext}. The regression branch is simply deployed as $\mathtt{Conv}\rightarrow\mathtt{GroupNorm}\rightarrow\mathtt{ReLU}\rightarrow\mathtt{Conv}$, with the last convolution outputting 6 channels. Code and models will be released.

\noindent\textbf{Model Training.} We train our \Method with a batch size of 4 using AdamW~\cite{AdamW} optimizer. Following~\cite{yucontext}, the cosine annealing schedule is adopted, with the first $5\%$ iterations of warm-up, maximum learning rate of $3\times 10^{-4}$, weight decay of $0.01$ and $\beta_{1} = 0.9$, $\beta_{2} =0.99$, The experiments are conducted on 2 NVIDIA A100 GPUs (40G) with two samples for each GPU. The final prediction has dimensions of $128 \times 128 \times 16$, which is upsampled to $256 \times 256 \times 32$ through trilinear interpolation to align with the ground truth. We use the VoxNT trick to remove the wrong labels as Eq.~\ref{eq:gt_refine}, with the scale threshold set to 30 for each axis. The efficiency experiments are conducted on a NVIDIA 4090 (commercial GPU), considering the more practical deployment property. The training requires around 19 GB of memory per sample, with about 9.0 training hours for SemanticKITTI and 18 hours for SSCBench-KITTI360, which is very friendly for the research groups with commercial GPUs. 

\subsection{Algorithmic details of the VoxNT Trick}
We present the detailed process of our VoxNT Trick in Algorithm~\ref{alg:free} with PyTorch-style pseudo code. By calling the function $\mathrm{compute\_all\_direction\_distances()}$, this algorithm aims to generate the ground-truth of 4D offset field $\hat{\Delta}$ only using the per-voxel class labels \({Y} \in \mathbb{N}^{X\times Y\times Z}\) ($\mathrm{gt\_occ}$). In $Y$, each item satisfies \(0 < {Y}_{i,j,k} < K\) with \(K\) representing the number of classes. 

In brief, for every scanning direction \(\mathbf{d} \in \{x^+, x^-, y^+, y^-, z^+, z^-\}\), we begin by initializing a zero-valued matrix \(R \in \mathbb{N}^{X\times Y\times Z}\). Then, as we iteratively traverse the voxels along the chosen direction \(d\), we compare consecutive voxels. Given the two adjacent voxels \(\mathbf{V}_{i}\) and \(\mathbf{V}_{i+1}\), suppose the class label of them matches with each other: ${Y}_{i} = {Y}_{i+1}$ considering the scanning direction, the corresponding entry in \(R\) is incremented by one; otherwise, the accumulation halts once a class change is detected, revealing approaching the instance boundary. 

This trick efficiently captures the spatial continuity of object instances and provides reliable ground truth offsets for regression. Considering the symmetrical property between the two scanning directions (forward and backward), we deploy a flip operation along the selected axis for efficient implementation. Based on this, we can generate the 4D tensor saving these relative distances in 6 directions, where each dimension is then normalized into $[0,1]$ divided by the volume size $X,Y,Z$ accordingly. Thus, we can obtain a normalized offset field $\hat{\Delta}$ ground-truth from semantic labels.

\textbf{Implementing \Method with LiDAR Input.} \Method is designed on the lifted 3D volumes, which can be effortlessly transferred to a LiDAR-based pipeline. To further analyze this flexibility, we deploy a LiDAR-based \Method, denoted as \textbf{\Method-L}. We achieve this by replacing the 2D-to-3D lifting with a simple point cloud encoder. This can directly generate the 3D feature volume $\Voxel$ with 3D ResNet-50, using LiDAR point cloud as input. The 2D image encoder, depth estimator, and view transformation are removed. \emph{All the implementations only use a single model without multi-frame distillation and only a single frame input without temporal information,} which will be open-sourced.

\definecolor{codeblue}{rgb}{0.25,0.5,0.5}
\definecolor{codekw}{rgb}{0.85, 0.18, 0.50}

\definecolor{codesign}{RGB}{0, 0, 255}
\definecolor{codefunc}{rgb}{0.85, 0.18, 0.50}

\lstdefinelanguage{PythonFuncColor}{
  language=Python,
  keywordstyle=\color{blue}\bfseries,
  commentstyle=\color{codeblue},  
  stringstyle=\color{orange},
  showstringspaces=false,
  basicstyle=\ttfamily\small,
  literate=
    {*}{{\color{codesign}* }}{1}
    {-}{{\color{codesign}- }}{1}
    {+}{{\color{codesign}+ }}{1}
    {dataloader}{{\color{codefunc}dataloader}}{1}
    {sample_t_r}{{\color{codefunc}sample\_t\_r}}{1}
    {randn}{{\color{codefunc}randn}}{1}
    {randn_like}{{\color{codefunc}randn\_like}}{1}
    {jvp}{{\color{codefunc}jvp}}{1}
    {stopgrad}{{\color{codefunc}stopgrad}}{1}
    {metric}{{\color{codefunc}metric}}{1}
}

\lstset{
  language=PythonFuncColor,
  backgroundcolor=\color{white},
  basicstyle=\fontsize{9pt}{9.9pt}\ttfamily\selectfont,
  columns=fullflexible,
  breaklines=true,
  captionpos=b,
}

\begin{algorithm}[t]
\caption{PyTorch Style Pseudocode of Voxel-to-Instance (\Trick) Trick.}
\label{alg:free}
\begin{lstlisting}
def compute_all_direction_distances(gt_occ):
    B, X, Y, Z = gt_occ.shape

    dist_x_pos = run_length_along_dim(gt_occ, 1, "positive")
    dist_x_neg = run_length_along_dim(gt_occ, 1, "negative")
    dist_y_pos = run_length_along_dim(gt_occ, 2, "positive")
    dist_y_neg = run_length_along_dim(gt_occ, 2, "negative")
    dist_z_pos = run_length_along_dim(gt_occ, 3, "positive")
    dist_z_neg = run_length_along_dim(gt_occ, 3, "negative")
    
    # Stack into shape (B, 6, X, Y, Z)
    distances = torch.stack([
        dist_x_pos, dist_x_neg,
        dist_y_pos, dist_y_neg,
        dist_z_pos, dist_z_neg
    ], dim=1)
    return distances
    
def run_length_along_dim(t, dim, direction):
    if direction == "positive":
        return run_length_positive(t, dim)
    else:
        # flip, compute positive, then flip back
        tf = torch.flip(t, dims=(dim,))
        of = run_length_positive(tf, dim)
        return torch.flip(of, dims=(dim,))

def run_length_positive(t, dim):
    shape = t.shape
    L = shape[dim]
    out = torch.empty_like(t, dtype=torch.int32)
    
    # Initialize the last slice along dim to 1
    idx_last = [slice(None)] * len(shape)
    idx_last[dim] = -1
    out[tuple(idx_last)] = torch.tensor(1, dtype=torch.int32, device=t.device)
    
    for i in range(L - 2, -1, -1):
        idx      = [slice(None)] * len(shape); idx[dim] = i
        idx_next = [slice(None)] * len(shape); idx_next[dim] = i + 1
        
        current  = t[tuple(idx)]
        nxt      = t[tuple(idx_next)]
        
        # Check whether the next voxel is in the same class
        same     = (current == nxt)
        out_next = out[tuple(idx_next)]
        
        out_val = torch.where(
            same,
            out_next + 1,
            torch.tensor(1, dtype=torch.int32, device=t.device)
        )
        out[tuple(idx)] = out_val
        
    return out

\end{lstlisting}
\end{algorithm}

\textbf{}

\section{Discussion}\label{sec:discussion}

\subsection{Difference with Related Works}

\noindent\textbf{Occupancy prediction assisted with 3D object detection.} Some existing works have attempted to assist the voxel prediction with 3D object detection~\cite{ming2025inverse++,yang2024daocc}. Although they consider the instance-level representation, our \Method is essentially different from these works in the following aspects. 

\textbf{(1) Extra-label Free.} Different from these works that require extra instance-level bounding box labels, \Method does not need any handcrafted bounding box labels for the object instances thanks to the proposed \Trick trick, which significantly reduces the labeling labor with more practical usage. 

\textbf{(2) Extension to Image and LiDAR based Pipelines.} Only using occupancy labels, \Method can be extended to both camera-based and LiDAR-based settings in a unified pipeline, achieving state-of-the-art results on both benchmarks. This differs from most existing works, which are tailored only for a single type of input modality. Hence, our method has more substantial flexibility and transferability.

\textbf{(3) Flexible Definition of Object Instances.} These works~\cite{ming2025inverse++,yang2024daocc} highly rely on the specific definition of instances in the detection datasets, which usually only considers some primary objects like cars and humans. However, extending to a more profound class set, such as buildings, is challenging and almost impossible, making this instance-level perception un-extensible. Differently, our \Trick can handle the instances in arbitrary classes, providing an extremely flexible definition for object instances. For example, we can discover the length, width, and height of buildings, vegetation, etc (see Fig.~\ref{fig:offset_distribution}), which is highly valuable for autonomous navigation.

\textbf{(4) Technical Difference.} Existing works~\cite{ming2025inverse++,yang2024daocc} introduce a separate 3D‐object‐detection branch to support occupancy prediction. In contrast, we consolidate both tasks into a single, detection‐driven formulation with an ultra-lightweight design, which can avoid the potential conflict between occupancy prediction and 3D object detection. Our feature decoupled designs also avoided the task-misalignment issue, which is ignored in existing works. The proposed dense‐regression design, i.e., each voxel regresses the distance to the instance boundary, is a different paradigm for instance-level perception, which is implemented purely with convolutional layers, eliminating the complexity and computational overhead of object-query–based attention.

\subsection{Broader Impacts}

\noindent\textbf{Impacts to the Broader Occupancy Community.} Our \Trick and \Method can be directly used in any voxel-included settings, such as multi-view occupancy prediction, as it is based on a similar voxel representation. By freely transferring semantic voxel labels into instance-level offset labels, this work may also inspire the occupancy community to reconsider the instance-level perception (e.g., instance and panoptic segmentation) achieved using only semantic labels, potentially contributing to training scale-up with practical label usage. 

\noindent\textbf{Impacts to the 3D Point Cloud Community.} As voxelization is a key procedure in point-cloud understanding, our method, fully based on voxel representation, has great potential for the point cloud community. We achieve state-of-the-art results on the LiDAR-based SemanticKITTI benchmark, which justifies our potential impact on the point cloud community.

\noindent\textbf{Impacts to the 3D Object Detection Community.} Our \Trick bridges the gap between dense occupancy and sparse 3D object detection. It converts semantically rich voxel-wise labels (e.g., car, building, vegetation) into instance-level offset labels that capture precise object extents and class-aware geometric priors. By integrating these instance labels into the occupancy, 3D detectors can revisit their image-driven origins, i.e., detecting objects directly from 2D inputs, while evolving toward a voxel-based instance detection paradigm that more faithfully reflects the physical 3D world.

\subsection{Limitations and Future Work}

\noindent\textbf{Generate Sparse 3D Bounding Box Visualization.} The current \Method uses instance-level supervision to guide the dense occupancy prediction, which can not directly generate spare 3D bounding boxes. The reason is that in dense detection~\cite{tian2019fcos,lin2017focal,law2018cornernet}, the Non-Maximum Suppression (NMS) is required to remove low-quality boxes for final visualizations. However, there is a gap between the 2D pixels and 3D voxels. To solve this, we will develop a 3D NMS algorithm tailored for voxels, bridging the gap between the voxel-level occupancy prediction and instance-level 3D object detection.

\begin{figure}[t]
\centering\includegraphics[width=0.9\linewidth]{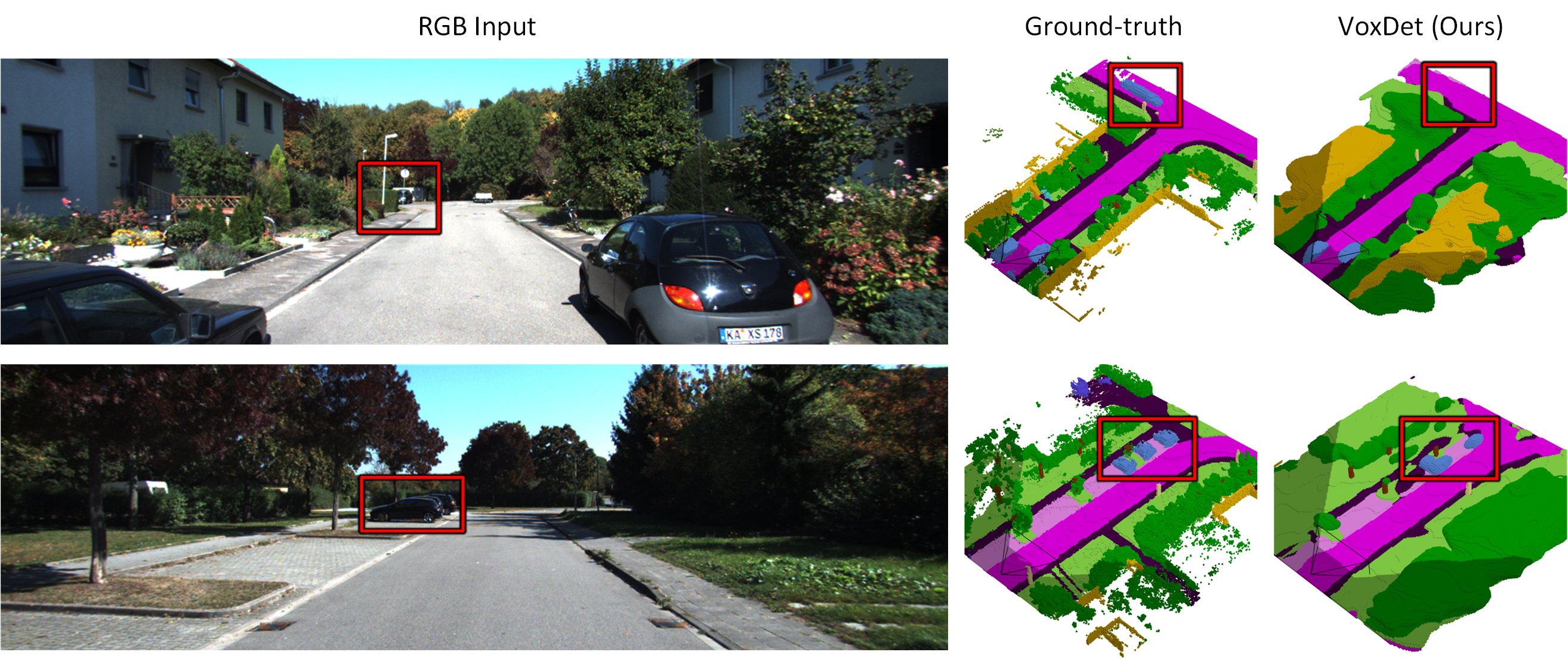}
\caption{Failure cases of the proposed \Method. It is difficult for our method to correctly detect objects at extremely far distances satisfactorily due to limited semantic cues on the RGB input images.}
\label{fig:fail}
\end{figure}

\noindent\textbf{Lack the Usage of Extra Temporal Information.} The current \Method is based on single-frame input. While state-of-the-art, the performance can be further improved with extra temporal information~\cite{li2024hierarchical,VoxFormer}, which can correct the cross-frame inconsistency. This will be our future work.

\noindent\textbf{Implementation on Multi-View Pipelines.} The current \Method is evaluated with a single camera input, which may limit its application scope. Our method can be transferred to the multi-view settings effortlessly, as it is deployed on the 3D feature volume, which is also the representation for multi-view occupancy prediction. This will be our future work.

\begin{figure}[t]
\centering\includegraphics[width=0.95\linewidth]{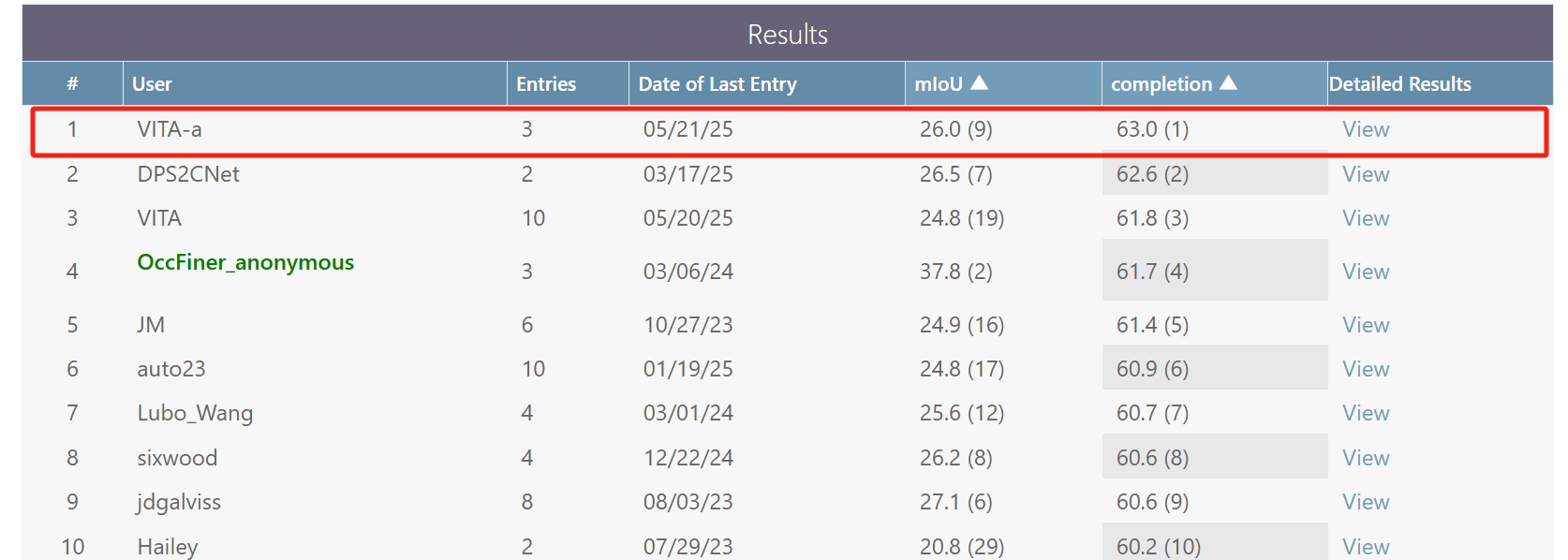}
\caption{Online leaderboard of SemanticKITTI hidden test set.}
\label{fig:leaderboard}
\end{figure}

\begin{figure}[t]
\centering\includegraphics[width=1.0\linewidth]{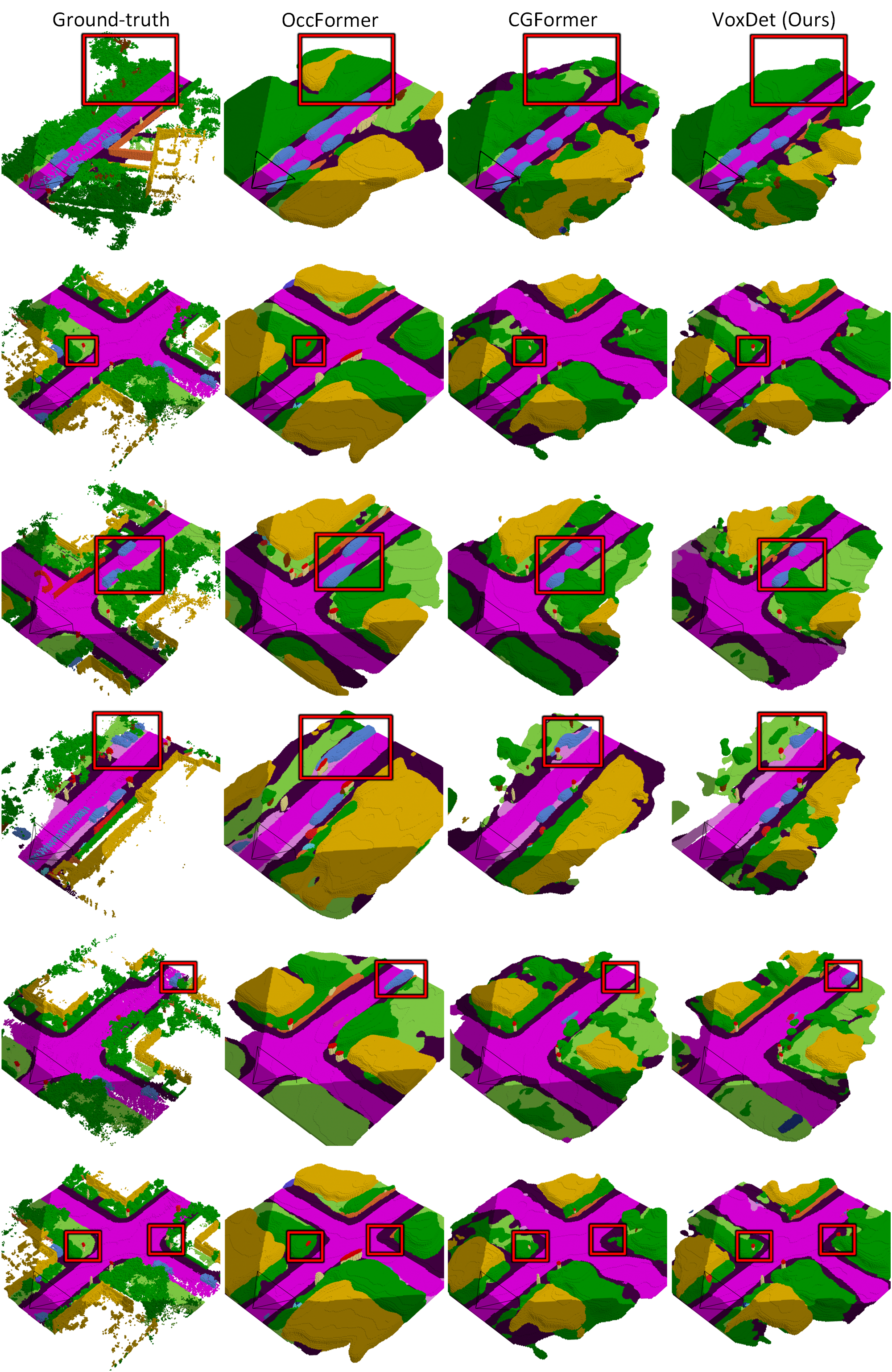}
\caption{More qualitative comparisons on SemanticKITTI validation set.}
\label{fig:more_qual_1}
\end{figure}

\begin{figure}[t]
\centering\includegraphics[width=1.0\linewidth]{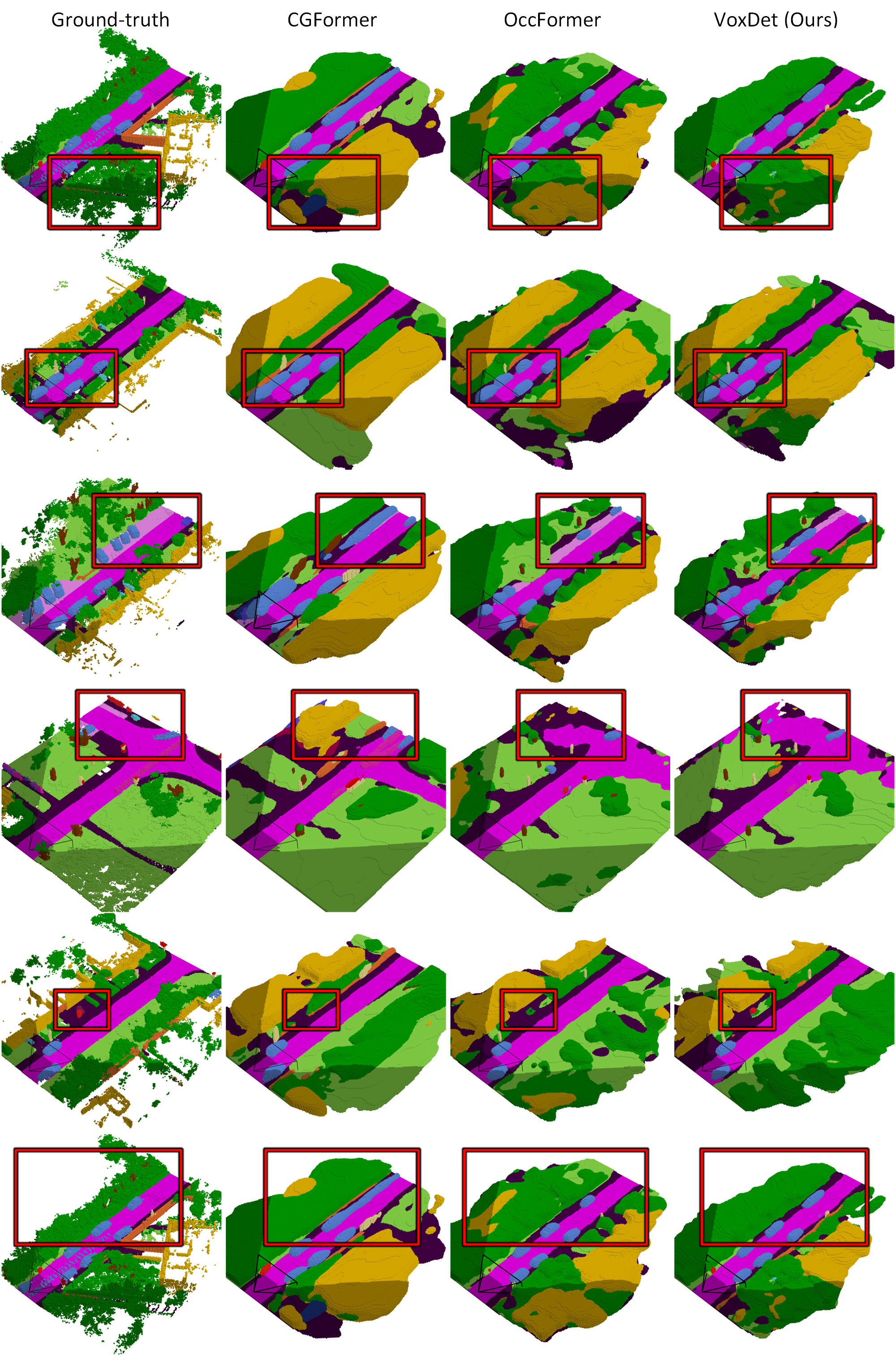}
\caption{More qualitative comparisons on SemanticKITTI validation set.}
\label{fig:more_qual_2}
\end{figure}

\noindent\textbf{Rely on the Depth Prediction Accuracy.} Although \Method achieves state-of-the-art performance, we empirically find that it will also suffer from some false-negative predictions of the objects at a long distance (see App.~\ref{sec:qual_exp}). The sub-optimal depth estimation may be the reason. In the main paper, we also find that replacing the stereo depth with monocular depth leads to some performance drop, highlighting the reliance on depth accuracy. Therefore, improving the depth models in the future may further enhance the performance and serve as our future work.

\subsection{Ethical Claims}

Our \Method uses only publicly available datasets and produces 3D semantic volumes without retaining any personally identifiable information. We do not foresee any significant negative impacts: the method does not enable individual tracking or intrusive surveillance, poses no additional privacy or safety risks beyond standard camera perception systems, and does not rely on sensitive demographic attributes. By focusing on high‐level scene understanding for benign applications such as autonomous navigation, our approach raises no known ethical, fairness, or regulatory concerns.

\section{Additional Qualitative Results}\label{sec:qual_exp}

\subsection{Failure Cases}
In Fig.~\ref{fig:fail}, we present some failure cases generated by the proposed \Method. We observe that our method fails to give correct detection for the object at extremely far distances (top sample), especially when objects of similar color are highly overlapped in the 2D images (bottom sample). This may be caused by the minimal visual cues on the 2D images, limiting the capacity. Some potential solutions may be (1) deploying more powerful visual feature extractors tailored for high-quality dense perception; (2) introducing extra modality, such as LiDAR, to enhance the information in the far distance; (3) developing tailored algorithms refining the features of objects in the far distance.

\subsection{More Comparison with Other Methods}

In Fig.~\ref{fig:more_qual_1} and \ref{fig:more_qual_2}, we provide additional qualitative comparisons against the state-of-the-art methods CGFormer \cite{yucontext} and OccFormer \cite{OccFormer}. Our approach exhibits visually superior instance-level completeness, greater environmental consistency, and enhanced semantic perception.

\subsection{Online Leaderboard}

Fig.~\ref{fig:leaderboard} illustrates the online leaderboard of SemanticKITTI hidden test set. Our method achieves the best 63.0 IoU (completion) only using the single-frame LiDAR input, showing the effectiveness.

\clearpage
\bibliography{references}

\begin{thebibliography}{10}\itemsep=-1pt

\bibitem{alahi2016social}
Alexandre Alahi, Kratarth Goel, Vignesh Ramanathan, Alexandre Robicquet, Li Fei-Fei, and Silvio Savarese.
\newblock Social lstm: Human trajectory prediction in crowded spaces.
\newblock In {\em Proceedings of the IEEE conference on computer vision and pattern recognition}, pages 961--971, 2016.

\bibitem{bae2024three}
Jongseong Bae, Junwoo Ha, and Ha~Young Kim.
\newblock Three cars approaching within 100m! enhancing distant geometry by tri-axis voxel scanning for camera-based semantic scene completion.
\newblock {\em arXiv preprint arXiv:2411.16129}, 2024.

\bibitem{SemanticKITTI}
Jens Behley, Martin Garbade, Andres Milioto, Jan Quenzel, Sven Behnke, Cyrill Stachniss, and J{\"{u}}rgen Gall.
\newblock Semantickitti: {A} dataset for semantic scene understanding of lidar sequences.
\newblock In {\em Proceedings of the IEEE/CVF International Conference on Computer Vision}, pages 9297--9307, 2019.

\bibitem{behley2019semantickitti}
Jens Behley, Martin Garbade, Andres Milioto, Jan Quenzel, Sven Behnke, Cyrill Stachniss, and Jurgen Gall.
\newblock Semantickitti: A dataset for semantic scene understanding of lidar sequences.
\newblock In {\em Proceedings of the IEEE/CVF international conference on computer vision}, pages 9297--9307, 2019.

\bibitem{Adabins}
Shariq~Farooq Bhat, Ibraheem Alhashim, and Peter Wonka.
\newblock Adabins: Depth estimation using adaptive bins.
\newblock In {\em Proceedings of the IEEE/CVF Conference on Computer Vision and Pattern Recognition}, pages 4009--4018, 2021.

\bibitem{MonoScene}
Anh{-}Quan Cao and Raoul de Charette.
\newblock Monoscene: Monocular 3d semantic scene completion.
\newblock In {\em Proceedings of the IEEE/CVF Conference on Computer Vision and Pattern Recognition}, pages 3981--3991, 2022.

\bibitem{cao2024pasco}
Anh-Quan Cao, Angela Dai, and Raoul De~Charette.
\newblock Pasco: Urban 3d panoptic scene completion with uncertainty awareness.
\newblock In {\em Proceedings of the IEEE/CVF Conference on Computer Vision and Pattern Recognition}, pages 14554--14564, 2024.

\bibitem{DETR}
Nicolas Carion, Francisco Massa, Gabriel Synnaeve, Nicolas Usunier, Alexander Kirillov, and Sergey Zagoruyko.
\newblock End-to-end object detection with transformers.
\newblock In {\em Proceedings of the European Conference on Computer Vision}, pages 213--229, 2020.

\bibitem{chambon2025gaussrender}
Lo{\~A}{\NG}ck Chambon, Eloi Zablocki, Alexandre Boulch, Micka{\~A}l Chen, and Matthieu Cord.
\newblock Gaussrender: Learning 3d occupancy with gaussian rendering.
\newblock {\em arXiv preprint arXiv:2502.05040}, 2025.

\bibitem{chen2018domain}
Yuhua Chen, Wen Li, Christos Sakaridis, Dengxin Dai, and Luc Van~Gool.
\newblock Domain adaptive faster r-cnn for object detection in the wild.
\newblock In {\em Proceedings of the IEEE conference on computer vision and pattern recognition}, pages 3339--3348, 2018.

\bibitem{chen2021disentangle}
Zehui Chen, Chenhongyi Yang, Qiaofei Li, Feng Zhao, Zheng-Jun Zha, and Feng Wu.
\newblock Disentangle your dense object detector.
\newblock In {\em Proceedings of the 29th ACM international conference on multimedia}, pages 4939--4948, 2021.

\bibitem{cciccek20163d}
{\"O}zg{\"u}n {\c{C}}i{\c{c}}ek, Ahmed Abdulkadir, Soeren~S Lienkamp, Thomas Brox, and Olaf Ronneberger.
\newblock 3d u-net: learning dense volumetric segmentation from sparse annotation.
\newblock In {\em Medical Image Computing and Computer-Assisted Intervention--MICCAI 2016: 19th International Conference, Athens, Greece, October 17-21, 2016, Proceedings, Part II 19}, pages 424--432. Springer, 2016.

\bibitem{dai2017deformable}
Jifeng Dai, Haozhi Qi, Yuwen Xiong, Yi Li, Guodong Zhang, Han Hu, and Yichen Wei.
\newblock Deformable convolutional networks.
\newblock In {\em Proceedings of the IEEE international conference on computer vision}, pages 764--773, 2017.

\bibitem{duan2025worldscore}
Haoyi Duan, Hong-Xing Yu, Sirui Chen, Li Fei-Fei, and Jiajun Wu.
\newblock Worldscore: A unified evaluation benchmark for world generation.
\newblock {\em arXiv preprint arXiv:2504.00983}, 2025.

\bibitem{faugeras1993three}
Olivier Faugeras.
\newblock {\em Three-dimensional computer vision: a geometric viewpoint}.
\newblock MIT press, 1993.

\bibitem{garbade2019two}
Martin Garbade, Yueh-Tung Chen, Johann Sawatzky, and Juergen Gall.
\newblock Two stream 3d semantic scene completion.
\newblock In {\em Proceedings of the IEEE/CVF Conference on Computer Vision and Pattern Recognition Workshops}, pages 0--0, 2019.

\bibitem{geiger2012we}
Andreas Geiger, Philip Lenz, and Raquel Urtasun.
\newblock Are we ready for autonomous driving? the kitti vision benchmark suite.
\newblock In {\em 2012 IEEE conference on computer vision and pattern recognition}, pages 3354--3361. IEEE, 2012.

\bibitem{girshick2015fast}
Ross Girshick.
\newblock Fast r-cnn.
\newblock In {\em Proceedings of the IEEE international conference on computer vision}, pages 1440--1448, 2015.

\bibitem{guo2025sgformer}
Xiyue Guo, Jiarui Hu, Junjie Hu, Hujun Bao, and Guofeng Zhang.
\newblock Sgformer: Satellite-ground fusion for 3d semantic scene completion.
\newblock In {\em Proceedings of the IEEE/CVF international conference on computer vision}, 2025.

\bibitem{MAE}
Kaiming He, Xinlei Chen, Saining Xie, Yanghao Li, Piotr Doll{\'a}r, and Ross Girshick.
\newblock Masked autoencoders are scalable vision learners.
\newblock In {\em Proceedings of the IEEE/CVF Conference on Computer Vision and Pattern Recognition}, pages 16000--16009, 2022.

\bibitem{ResNet}
Kaiming He, Xiangyu Zhang, Shaoqing Ren, and Jian Sun.
\newblock Deep residual learning for image recognition.
\newblock In {\em Proceedings of the IEEE/CVF conference on computer vision and pattern recognition}, pages 770--778, 2016.

\bibitem{TPVFormer}
Yuanhui Huang, Wenzhao Zheng, Yunpeng Zhang, Jie Zhou, and Jiwen Lu.
\newblock Tri-perspective view for vision-based 3d semantic occupancy prediction.
\newblock In {\em Proceedings of the IEEE/CVF Conference on Computer Vision and Pattern Recognition}, pages 9223--9232, 2023.

\bibitem{Symphonize}
Haoyi Jiang, Tianheng Cheng, Naiyu Gao, Haoyang Zhang, Wenyu Liu, and Xinggang Wang.
\newblock Symphonize 3d semantic scene completion with contextual instance queries.
\newblock {\em Proceedings of the IEEE/CVF Conference on Computer Vision and Pattern Recognition}, 2024.

\bibitem{law2018cornernet}
Hei Law and Jia Deng.
\newblock Cornernet: Detecting objects as paired keypoints.
\newblock In {\em Proceedings of the European conference on computer vision (ECCV)}, pages 734--750, 2018.

\bibitem{li2024hierarchical}
Bohan Li, Jiajun Deng, Wenyao Zhang, Zhujin Liang, Dalong Du, Xin Jin, and Wenjun Zeng.
\newblock Hierarchical temporal context learning for camera-based semantic scene completion.
\newblock In {\em European Conference on Computer Vision}, 2024.

\bibitem{StereoScene}
Bohan Li, Yasheng Sun, Xin Jin, Wenjun Zeng, Zheng Zhu, Xiaoefeng Wang, Yunpeng Zhang, James Okae, Hang Xiao, and Dalong Du.
\newblock Stereoscene: Bev-assisted stereo matching empowers 3d semantic scene completion.
\newblock {\em arXiv preprint arXiv:2303.13959}, 2023.

\bibitem{li2025instantsplamp}
Chenxin Li, Hengyu Liu, Zhiwen Fan, Wuyang Li, Yifan Liu, Panwang Pan, and Yixuan Yuan.
\newblock Instantsplamp: Fast and generalizable stenography framework for generative gaussian splatting.
\newblock In {\em The Thirteenth International Conference on Learning Representations}, 2025.

\bibitem{li2025u}
Chenxin Li, Xinyu Liu, Wuyang Li, Cheng Wang, Hengyu Liu, Yifan Liu, Zhen Chen, and Yixuan Yuan.
\newblock U-kan makes strong backbone for medical image segmentation and generation.
\newblock In {\em Proceedings of the AAAI Conference on Artificial Intelligence}, volume~39, pages 4652--4660, 2025.

\bibitem{li2024occmamba}
Heng Li, Yuenan Hou, Xiaohan Xing, Yuexin Ma, Xiao Sun, and Yanyong Zhang.
\newblock Occmamba: Semantic occupancy prediction with state space models.
\newblock {\em arXiv preprint arXiv:2408.09859}, 2024.

\bibitem{li2021htd}
Wuyang Li, Zhen Chen, Baopu Li, Dingwen Zhang, and Yixuan Yuan.
\newblock Htd: Heterogeneous task decoupling for two-stage object detection.
\newblock {\em IEEE Transactions on Image Processing}, 30:9456--9469, 2021.

\bibitem{li2022scan}
Wuyang Li, Xinyu Liu, Xiwen Yao, and Yixuan Yuan.
\newblock Scan: Cross domain object detection with semantic conditioned adaptation.
\newblock In {\em Proceedings of the AAAI Conference on Artificial Intelligence}, 2022.

\bibitem{li2022sigma}
Wuyang Li, Xinyu Liu, and Yixuan Yuan.
\newblock Sigma: Semantic-complete graph matching for domain adaptive object detection.
\newblock In {\em Proceedings of the IEEE/CVF conference on computer vision and pattern recognition}, pages 5291--5300, 2022.

\bibitem{li2022unifying}
Yanwei Li, Yilun Chen, Xiaojuan Qi, Zeming Li, Jian Sun, and Jiaya Jia.
\newblock Unifying voxel-based representation with transformer for 3d object detection.
\newblock {\em Advances in Neural Information Processing Systems}, 35:18442--18455, 2022.

\bibitem{li2024sscbenchlargescale3dsemantic}
Yiming Li, Sihang Li, Xinhao Liu, Moonjun Gong, Kenan Li, Nuo Chen, Zijun Wang, Zhiheng Li, Tao Jiang, Fisher Yu, Yue Wang, Hang Zhao, Zhiding Yu, and Chen Feng.
\newblock Sscbench: A large-scale 3d semantic scene completion benchmark for autonomous driving, 2024.

\bibitem{VoxFormer}
Yiming Li, Zhiding Yu, Christopher~B. Choy, Chaowei Xiao, Jos{\'{e}}~M. {\'{A}}lvarez, Sanja Fidler, Chen Feng, and Anima Anandkumar.
\newblock Voxformer: Sparse voxel transformer for camera-based 3d semantic scene completion.
\newblock In {\em Proceedings of the IEEE/CVF Conference on Computer Vision and Pattern Recognition}, pages 9087--9098, 2023.

\bibitem{liang2025skip}
Li Liang, Naveed Akhtar, Jordan Vice, Xiangrui Kong, and Ajmal~Saeed Mian.
\newblock Skip mamba diffusion for monocular 3d semantic scene completion.
\newblock {\em arXiv preprint arXiv:2501.07260}, 2025.

\bibitem{KITTI360}
Yiyi Liao, Jun Xie, and Andreas Geiger.
\newblock Kitti-360: A novel dataset and benchmarks for urban scene understanding in 2d and 3d.
\newblock {\em IEEE Transactions on Pattern Analysis and Machine Intelligence}, pages 3292--3310, 2022.

\bibitem{liao2022kitti}
Yiyi Liao, Jun Xie, and Andreas Geiger.
\newblock Kitti-360: A novel dataset and benchmarks for urban scene understanding in 2d and 3d.
\newblock {\em IEEE Transactions on Pattern Analysis and Machine Intelligence}, 45(3):3292--3310, 2022.

\bibitem{FPN}
Tsung-Yi Lin, Piotr Doll{\'a}r, Ross Girshick, Kaiming He, Bharath Hariharan, and Serge Belongie.
\newblock Feature pyramid networks for object detection.
\newblock In {\em Proceedings of the IEEE/CVF Conference on Computer Vision and Pattern Recognition}, pages 2117--2125, 2017.

\bibitem{lin2017focal}
Tsung-Yi Lin, Priya Goyal, Ross Girshick, Kaiming He, and Piotr Doll{\'a}r.
\newblock Focal loss for dense object detection.
\newblock In {\em Proceedings of the IEEE international conference on computer vision}, pages 2980--2988, 2017.

\bibitem{liu2024lgs}
Hengyu Liu, Yifan Liu, Chenxin Li, Wuyang Li, and Yixuan Yuan.
\newblock Lgs: A light-weight 4d gaussian splatting for efficient surgical scene reconstruction.
\newblock In {\em International Conference on Medical Image Computing and Computer-Assisted Intervention}, pages 660--670. Springer, 2024.

\bibitem{liu2025x}
Yifan Liu, Wuyang Li, Weihao Yu, Chenxin Li, Alexandre Alahi, Max Meng, and Yixuan Yuan.
\newblock X-grm: Large gaussian reconstruction model for sparse-view x-rays to computed tomography.
\newblock {\em arXiv preprint arXiv:2505.15235}, 2025.

\bibitem{AdamW}
Ilya Loshchilov and Frank Hutter.
\newblock Decoupled weight decay regularization.
\newblock {\em arXiv preprint arXiv:1711.05101}, 2017.

\bibitem{rtDETR}
Wenyu Lv, Shangliang Xu, Yian Zhao, Guanzhong Wang, Jinman Wei, Cheng Cui, Yuning Du, Qingqing Dang, and Yi Liu.
\newblock Detrs beat yolos on real-time object detection.
\newblock {\em arXiv preprint arXiv:2304.08069}, 2023.

\bibitem{SGN}
Jianbiao Mei, Yu Yang, Mengmeng Wang, Junyu Zhu, Xiangrui Zhao, Jongwon Ra, Laijian Li, and Yong Liu.
\newblock Camera-based 3d semantic scene completion with sparse guidance network.
\newblock {\em arXiv preprint arXiv:2312.05752}, 2023.

\bibitem{ming2025inverse++}
Zhenxing Ming, Julie~Stephany Berrio, Mao Shan, and Stewart Worrall.
\newblock Inverse++: Vision-centric 3d semantic occupancy prediction assisted with 3d object detection.
\newblock {\em arXiv preprint arXiv:2504.04732}, 2025.

\bibitem{lss}
Jonah Philion and Sanja Fidler.
\newblock Lift, splat, shoot: Encoding images from arbitrary camera rigs by implicitly unprojecting to 3d.
\newblock In {\em Proceedings of the European Conference on Computer Vision}, pages 194--210, 2020.

\bibitem{qiu2020borderdet}
Han Qiu, Yuchen Ma, Zeming Li, Songtao Liu, and Jian Sun.
\newblock Borderdet: Border feature for dense object detection.
\newblock In {\em European Conference on Computer Vision}, pages 549--564. Springer, 2020.

\bibitem{ren2016faster}
Shaoqing Ren, Kaiming He, Ross Girshick, and Jian Sun.
\newblock Faster r-cnn: Towards real-time object detection with region proposal networks.
\newblock {\em IEEE transactions on pattern analysis and machine intelligence}, 39(6):1137--1149, 2016.

\bibitem{rist2021semantic}
Christoph~B Rist, David Emmerichs, Markus Enzweiler, and Dariu~M Gavrila.
\newblock Semantic scene completion using local deep implicit functions on lidar data.
\newblock {\em IEEE transactions on pattern analysis and machine intelligence}, 44(10):7205--7218, 2021.

\bibitem{LMSCNet}
Luis Rold{\~{a}}o, Raoul de Charette, and Anne Verroust{-}Blondet.
\newblock Lmscnet: Lightweight multiscale 3d semantic completion.
\newblock In {\em Proceedings of the International Conference on 3D Vision}, pages 111--119, 2020.

\bibitem{roldao20223d}
Luis Roldao, Raoul De~Charette, and Anne Verroust-Blondet.
\newblock 3d semantic scene completion: A survey.
\newblock {\em International Journal of Computer Vision}, 130(8):1978--2005, 2022.

\bibitem{MobileStereoNet}
Faranak Shamsafar, Samuel Woerz, Rafia Rahim, and Andreas Zell.
\newblock Mobilestereonet: Towards lightweight deep networks for stereo matching.
\newblock In {\em Proceedings of the IEEE/CVF Winter Conference on Applications of Computer Vision}, pages 2417--2426, 2022.

\bibitem{song2020revisiting}
Guanglu Song, Yu Liu, and Xiaogang Wang.
\newblock Revisiting the sibling head in object detector.
\newblock In {\em Proceedings of the IEEE/CVF conference on computer vision and pattern recognition}, pages 11563--11572, 2020.

\bibitem{SSCNet}
Shuran Song, Fisher Yu, Andy Zeng, Angel~X. Chang, Manolis Savva, and Thomas~A. Funkhouser.
\newblock Semantic scene completion from a single depth image.
\newblock In {\em Proceedings of the IEEE/CVF Conference on Computer Vision and Pattern Recognition}, pages 190--198, 2017.

\bibitem{tian2019fcos}
Zhi Tian, Chunhua Shen, Hao Chen, and Tong He.
\newblock Fcos: Fully convolutional one-stage object detection.
\newblock In {\em Proceedings of the IEEE/CVF international conference on computer vision}, pages 9627--9636, 2019.

\bibitem{wang2024voxel}
Lubo Wang, Di Lin, Kairui Yang, Ruonan Liu, Qing Guo, Wuyuan Xie, Miaohui Wang, Lingyu Liang, Yi Wang, and Ping Li.
\newblock Voxel proposal network via multi-frame knowledge distillation for semantic scene completion.
\newblock {\em Advances in Neural Information Processing Systems}, 37:101096--101115, 2024.

\bibitem{wang2025vlscene}
Meng Wang, Huilong Pi, Ruihui Li, Yunchuan Qin, Zhuo Tang, and Kenli Li.
\newblock Vlscene: Vision-language guidance distillation for camera-based 3d semantic scene completion.
\newblock In {\em Proceedings of the AAAI Conference on Artificial Intelligence}, volume~39, pages 7808--7816, 2025.

\bibitem{wang2020fcos}
Ning Wang, Yang Gao, Hao Chen, Peng Wang, Zhi Tian, Chunhua Shen, and Yanning Zhang.
\newblock Nas-fcos: Fast neural architecture search for object detection.
\newblock In {\em proceedings of the IEEE/CVF conference on computer vision and pattern recognition}, pages 11943--11951, 2020.

\bibitem{wang2025l2cocc}
Ruoyu Wang, Yukai Ma, Yi Yao, Sheng Tao, Haoang Li, Zongzhi Zhu, Yong Liu, and Xingxing Zuo.
\newblock L2cocc: Lightweight camera-centric semantic scene completion via distillation of lidar model.
\newblock {\em arXiv preprint arXiv:2503.12369}, 2025.

\bibitem{HASSC}
Song Wang, Jiawei Yu, Wentong Li, Wenyu Liu, Xiaolu Liu, Junbo Chen, and Jianke Zhu.
\newblock Not all voxels are equal: Hardness-aware semantic scene completion with self-distillation.
\newblock {\em arXiv preprint arXiv:2404.11958}, 2024.

\bibitem{wang2021fcos3d}
Tai Wang, Xinge Zhu, Jiangmiao Pang, and Dahua Lin.
\newblock Fcos3d: Fully convolutional one-stage monocular 3d object detection.
\newblock In {\em Proceedings of the IEEE/CVF international conference on computer vision}, pages 913--922, 2021.

\bibitem{DETR3D}
Yue Wang, Vitor Guizilini, Tianyuan Zhang, Yilun Wang, Hang Zhao, , and Justin~M. Solomon.
\newblock Detr3d: 3d object detection from multi-view images via 3d-to-2d queries.
\newblock In {\em Conference on Robot Learning}, pages 180--191, 2021.

\bibitem{H2GFormer}
Yu Wang and Chao Tong.
\newblock H2gformer: Horizontal-to-global voxel transformer for 3d semantic scene completion.
\newblock In {\em Proceedings of the AAAI Conference on Artificial Intelligence}, pages 5722--5730, 2024.

\bibitem{surroundOcc}
Yi Wei, Linqing Zhao, Wenzhao Zheng, Zheng Zhu, Jie Zhou, and Jiwen Lu.
\newblock Surroundocc: Multi-camera 3d occupancy prediction for autonomous driving.
\newblock In {\em Proceedings of the IEEE/CVF International Conference on Computer Vision}, pages 21729--21740, 2023.

\bibitem{xia2023scpnet}
Zhaoyang Xia, Youquan Liu, Xin Li, Xinge Zhu, Yuexin Ma, Yikang Li, Yuenan Hou, and Yu Qiao.
\newblock Scpnet: Semantic scene completion on point cloud.
\newblock In {\em Proceedings of the IEEE/CVF conference on computer vision and pattern recognition}, pages 17642--17651, 2023.

\bibitem{IAMSSC}
Haihong Xiao, Hongbin Xu, Wenxiong Kang, and Yuqiong Li.
\newblock Instance-aware monocular 3d semantic scene completion.
\newblock {\em IEEE Transactions on Intelligent Transportation Systems}, 2024.

\bibitem{xu2022revisiting}
Dongli Xu, Jinhong Deng, and Wen Li.
\newblock Revisiting ap loss for dense object detection: Adaptive ranking pair selection.
\newblock In {\em Proceedings of the IEEE/CVF conference on computer vision and pattern recognition}, pages 14187--14196, 2022.

\bibitem{yan2021sparse}
Xu Yan, Jiantao Gao, Jie Li, Ruimao Zhang, Zhen Li, Rui Huang, and Shuguang Cui.
\newblock Sparse single sweep lidar point cloud segmentation via learning contextual shape priors from scene completion.
\newblock In {\em Proceedings of the AAAI conference on artificial intelligence}, pages 3101--3109, 2021.

\bibitem{yan2025event}
Zhiqiang Yan, Jianhao Jiao, Zhengxue Wang, and Gim~Hee Lee.
\newblock Event-driven dynamic scene depth completion.
\newblock {\em arXiv preprint arXiv:2505.13279}, 2025.

\bibitem{yan2025rignet++}
Zhiqiang Yan, Xiang Li, Le Hui, Zhenyu Zhang, Jun Li, and Jian Yang.
\newblock Rignet++: Semantic assisted repetitive image guided network for depth completion.
\newblock {\em International Journal of Computer Vision}, pages 1--23, 2025.

\bibitem{yan2024tri}
Zhiqiang Yan, Yuankai Lin, Kun Wang, Yupeng Zheng, Yufei Wang, Zhenyu Zhang, Jun Li, and Jian Yang.
\newblock Tri-perspective view decomposition for geometry-aware depth completion.
\newblock In {\em Proceedings of the IEEE/CVF Conference on Computer Vision and Pattern Recognition}, pages 4874--4884, 2024.

\bibitem{yan2022rignet}
Zhiqiang Yan, Kun Wang, Xiang Li, Zhenyu Zhang, Jun Li, and Jian Yang.
\newblock Rignet: Repetitive image guided network for depth completion.
\newblock In {\em European Conference on Computer Vision}, pages 214--230. Springer, 2022.

\bibitem{yang2022prediction}
Chenhongyi Yang, Mateusz Ochal, Amos Storkey, and Elliot~J Crowley.
\newblock Prediction-guided distillation for dense object detection.
\newblock In {\em European conference on computer vision}, pages 123--138. Springer, 2022.

\bibitem{yang2021semantic}
Xuemeng Yang, Hao Zou, Xin Kong, Tianxin Huang, Yong Liu, Wanlong Li, Feng Wen, and Hongbo Zhang.
\newblock Semantic segmentation-assisted scene completion for lidar point clouds.
\newblock In {\em 2021 IEEE/RSJ International Conference on Intelligent Robots and Systems (IROS)}, pages 3555--3562. IEEE, 2021.

\bibitem{yang2025concealgs}
Yifeng Yang, Hengyu Liu, Chenxin Li, Yining Sun, Wuyang Li, Yifan Liu, Yiyang Lin, Yixuan Yuan, and Nanyang Ye.
\newblock Concealgs: Concealing invisible copyright information in 3d gaussian splatting.
\newblock {\em arXiv preprint arXiv:2501.03605}, 2025.

\bibitem{yang2024daocc}
Zhen Yang, Yanpeng Dong, Heng Wang, Lichao Ma, Zijian Cui, Qi Liu, and Haoran Pei.
\newblock Daocc: 3d object detection assisted multi-sensor fusion for 3d occupancy prediction.
\newblock {\em arXiv preprint arXiv:2409.19972}, 2024.

\bibitem{DepthSSC}
Jiawei Yao and Jusheng Zhang.
\newblock Depthssc: Depth-spatial alignment and dynamic voxel resolution for monocular 3d semantic scene completion.
\newblock {\em arXiv preprint arXiv:2311.17084}, 2023.

\bibitem{yu2024language}
Zhu Yu, Bowen Pang, Lizhe Liu, Runmin Zhang, Qihao Peng, Maochun Luo, Sheng Yang, Mingxia Chen, Si-Yuan Cao, and Hui-Liang Shen.
\newblock Language driven occupancy prediction.
\newblock {\em arXiv preprint arXiv:2411.16072}, 2024.

\bibitem{yu2023aggregating}
Zhu Yu, Zehua Sheng, Zili Zhou, Lun Luo, Si-Yuan Cao, Hong Gu, Huaqi Zhang, and Hui-Liang Shen.
\newblock Aggregating feature point cloud for depth completion.
\newblock In {\em Proceedings of the IEEE/CVF international conference on computer vision}, pages 8732--8743, 2023.

\bibitem{PointDC}
Zhu Yu, Zehua Sheng, Zili Zhou, Lun Luo, Si-Yuan Cao, Hong Gu, Huaqi Zhang, and Hui-Liang Shen.
\newblock Aggregating feature point cloud for depth completion.
\newblock In {\em Proceedings of the IEEE/CVF International Conference on Computer Vision}, pages 8732--8743, 2023.

\bibitem{yucontext}
Zhu Yu, Runmin Zhang, Jiacheng Ying, Junchen Yu, Xiaohai Hu, Lun Luo, Si-Yuan Cao, and Hui-liang Shen.
\newblock Context and geometry aware voxel transformer for semantic scene completion.
\newblock In {\em The Thirty-eighth Annual Conference on Neural Information Processing Systems}, 2024.

\bibitem{zhang2021varifocalnet}
Haoyang Zhang, Ying Wang, Feras Dayoub, and Niko Sunderhauf.
\newblock Varifocalnet: An iou-aware dense object detector.
\newblock In {\em Proceedings of the IEEE/CVF conference on computer vision and pattern recognition}, pages 8514--8523, 2021.

\bibitem{zhang2020bridging}
Shifeng Zhang, Cheng Chi, Yongqiang Yao, Zhen Lei, and Stan~Z Li.
\newblock Bridging the gap between anchor-based and anchor-free detection via adaptive training sample selection.
\newblock In {\em Proceedings of the IEEE/CVF conference on computer vision and pattern recognition}, pages 9759--9768, 2020.

\bibitem{OccFormer}
Yunpeng Zhang, Zheng Zhu, and Dalong Du.
\newblock Occformer: Dual-path transformer for vision-based 3d semantic occupancy prediction.
\newblock {\em Proceedings of the IEEE/CVF Conference on Computer Vision and Pattern Recognition}, pages 9433--9443, 2023.

\bibitem{MonoOcc}
Yupeng Zheng, Xiang Li, Pengfei Li, Yuhang Zheng, Bu Jin, Chengliang Zhong, Xiaoxiao Long, Hao Zhao, and Qichao Zhang.
\newblock Monoocc: Digging into monocular semantic occupancy prediction.
\newblock {\em arXiv preprint arXiv:2403.08766}, 2024.

\bibitem{zheng2022localization}
Zhaohui Zheng, Rongguang Ye, Ping Wang, Dongwei Ren, Wangmeng Zuo, Qibin Hou, and Ming-Ming Cheng.
\newblock Localization distillation for dense object detection.
\newblock In {\em Proceedings of the IEEE/CVF conference on computer vision and pattern recognition}, pages 9407--9416, 2022.

\bibitem{zhou2019bottom}
Xingyi Zhou, Jiacheng Zhuo, and Philipp Krahenbuhl.
\newblock Bottom-up object detection by grouping extreme and center points.
\newblock In {\em Proceedings of the IEEE/CVF conference on computer vision and pattern recognition}, pages 850--859, 2019.

\bibitem{DeformableDetr}
Xizhou Zhu, Weijie Su, Lewei Lu, Bin Li, Xiaogang Wang, and Jifeng Dai.
\newblock Deformable detr: Deformable transformers for end-to-end object detection.
\newblock {\em arXiv preprint arXiv:2010.04159}, 2020.

\end{thebibliography}
\bibliographystyle{ieee_fullname}

\end{document}